\documentclass[journal,twoside]{IEEEtran}
\usepackage{booktabs}
\usepackage{multirow}
\usepackage{graphicx}
\usepackage{algorithmic}
\usepackage[linesnumbered,boxed,ruled,commentsnumbered]{algorithm2e}
\usepackage{amsmath,amssymb,amsfonts}
\usepackage{cite}
\newcommand{\RNum}[1]{\uppercase\expandafter{\romannumeral #1\relax}}
\usepackage{verbatim}
\usepackage{array}

\ifCLASSINFOpdf
\else
\fi
\begin{document}
%
\title{An Efficient Hardware Accelerator for Structured Sparse Convolutional Neural Networks on FPGAs}
%
%
%
\author{Chaoyang~Zhu,~\IEEEmembership{Student~Member,~IEEE,}\thanks{C. Zhu, S. Yang, Z. Zhu, K. Huang and H. Shen are with the College of Information Science and Electronic Engineering, Zhejiang University, Hangzhou, 310058, China (e\-mail: \{21760249,~huangkejie,~21931061,~21960370,~shen\_hb\}@zju.edu.cn).} Kejie~Huang,~\IEEEmembership{Senior~Member,~IEEE,} Shuyuan~Yang,~\IEEEmembership{Student~Member,~IEEE,} Ziqi~Zhu,~\IEEEmembership{Student~Member,~IEEE,} \\ Hejia~Zhang\thanks{H. Zhang is with the Department of Computer Science, University of Southern California, Los Angeles, CA 90089, USA (e\-mail: hejiazha@usc.edu)} and Haibin~Shen}


%
%


\markboth{This manuscirpt is for IEEE TRANSACTIONS ON VERY LARGE SCALE INTEGRATION (VLSI) SYSTEMS}%
{Zhu \MakeLowercase{\textit{et al.}}: An Efficient Hardware Accelerator for Structured Sparse Convolutional Neural Networks on FPGAs}
%



\maketitle

\begin{abstract}
Deep Convolutional Neural Networks (CNNs) have achieved state-of-the-art performance in a wide range of applications. However, deeper CNN models, which are usually computation consuming, are widely required for complex Artificial Intelligence (AI) tasks. Though recent research progress on network compression such as pruning has emerged as a promising direction to mitigate computational burden, existing accelerators are still prevented from completely utilizing the benefits of leveraging \textit{sparsity} owing to the \textit{irregularity} caused by pruning. On the other hand, Field-Programmable Gate Arrays (FPGAs) have been regarded as a promising hardware platform for CNN inference acceleration. However, most existing FPGA accelerators focus on dense CNN and cannot address the \textit{irregularity} problem. In this paper, we propose a sparse wise dataflow to skip the cycles of processing Multiply-and-Accumulates (MACs) with zero weights and exploit data statistics to minimize energy through zeros gating to avoid unnecessary computations. The proposed sparse wise dataflow leads to a low bandwidth requirement and a high data sharing. Then we design an FPGA accelerator containing a Vector Generator Module (VGM) which can match the index between sparse weights and input activations according to the proposed dataflow. Experimental results demonstrate that our implementation can achieve 987\,imag/s and 48\,imag/s performance for AlexNet and VGG-16 on Xilinx ZCU102, respectively, which provides 1.5$\times$ to 6.7$\times$ speedup and 2.0$\times$ to 6.2$\times$ energy-efficiency over previous CNN FPGA accelerators.
\end{abstract}


%
\IEEEpeerreviewmaketitle

\section{Introduction}
%
%
%
%
\IEEEPARstart{T}{he} remarkable performance improvement in various domains~\cite{he2015delving,girshick2014rich,redmon2016you} achieved by Convolutional Neural Networks (CNNs) like AlexNet~\cite{NIPS2012_4824} and VGG-16~\cite{simonyan2014very} comes at the computational and data cost which challenges both of on-chip storage and off-chip bandwidth in accelerator architecture design. Even though most of the operations in the CNN training and inference can be converted to matrix multiplication operations and be accelerated with modern Graphics Processing Units (GPUs), deploying CNNs on GPUs suffers from high power and area consumption. Customized accelerators have been regarded as a promising alternative which are more flexible for considering the performance requirements and energy constraints~\cite{dnnweaver2016,ma2018alamo,zeng2018framework,ma2017automatic,ma2018optimizing,podili2017fast,guo2017angel,li2016high,zhang2015optimizing,motamedi2016design,lu2017evaluating,ma2016scalable}.

Recently, the CNN pruning technique has been proved as an effective solution to reduce the computation and memory requirements of these models~\cite{han2015learning,han2015deep}. For example, Han et al.~\cite{han2015learning} pointed out that pruning can lead to more than 10$\times$ amount reduction of data with negligible accuracy loss. On the other hand, weight encoding including quantization and entropy coding has been proposed to further reduce the bitwidth of each weight, e.g., 4-bit per weight for AlexNet~\cite{han2015deep}. Unstructured pruning techniques like Deep Compression~\cite{han2015deep} have the weaknesses of imbalanced load and high irregularity. Therefore, structured pruning techniques~\cite{ding2017c, deng2018permdnn, zhou2018cambricon, zhang2018adam} were proposed which are more hardware friendly with a slightly lower compression ratio.

However, \textit{irregularity} caused by \textit{sparsity} prevents accelerators from fully leveraging the computation and data reduction. Exciting architectures on FPGAs for dense models are not efficient for sparse CNN models because a lot of weights are pruned so that most multiplication operations involve zero operands leading to low hardware efficiency~\cite{zhang2017improving,ma2017optimizing,guan2017fp,zeng2018framework,zhang2018accdnn,wei2017automated}. Sparse architecture on Field-Programmable Gate Arrays (FPGAs) has been investigated in recent years~\cite{han2017ese,lu2019efficient}. \cite{han2017ese} is designed for the Fully-Connected (FC) layers, which uses matrix-vector multiplication operations. In fact, the major operators in CNNs are convolution operations. Although the spatial convolution can be converted to matrix-vector multiplications, this will lead to a large memory footprint since the input feature map has to be copied multiple times when being flattened to a vector. \cite{lu2019efficient} proposes a dataflow that exploits element-matrix multiplication as the key operation. However, this design holds a low computation efficiency due to the imbalanced load of each Processing Engine (PE). Since this accelerator requires a large number of Look-Up-Table (LUT) to buffer the input activations for nonzero weights, the performance is bounded by the number of LUT on FPGA, which leads to an inefficient resource utilization.

To design an efficient FPGA accelerator for sparse CNN models, following challenges have to be tackled:

First of all, each output activation connects to several input activations through the sliding window (Kernel) in dense Convolutional (CONV) layers. The connection becomes irregular after pruning. Meanwhile, the sparse weights are encoded in compressed format, which results in extra coordinate computation to reconstruct the connection or to locate the output. So it is challenging to design a dataflow to address the \textit{irregularity}  whereas efficiently leverage the data and computation reduction and maintain the high parallelism of FPGA.


Second, FPGAs can only provide limited on-chip memory and off-chip bandwidth. Although sparse CNNs have significant data reduction, it is difficult to save all the weights into on-chip memory for complex CNNs like VGG-16. In addition, different CNN models have different sizes, which results in high variability in the number of operations. A rigid accelerator architecture for CNNs may not fully utilize the FPGA's limited resources for every CNN model.

To address both challenges, we only focus on structured pruning to reduce the irregularity of sparse weights. On this basis, a sparse wise dataflow is proposed 
to address the remaining irregularity of sparse weights. With proposed dataflow, we do not need extra coordinate computation to reconstruct the connection or to locate the output. Furthermore, we minimize energy through zeros gating to avoid unnecessary computations if the input activations equal zero.

\begin{figure}[t!]
	\centering
	\includegraphics[scale=0.5]{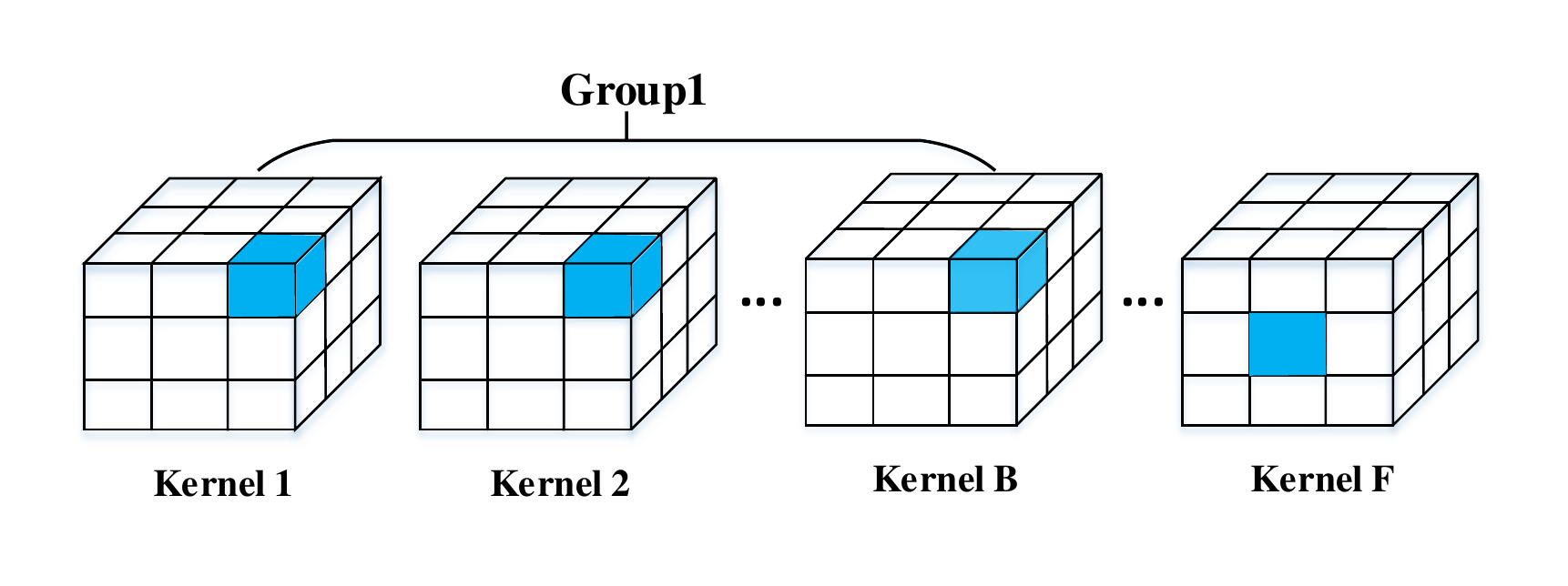}\\
	\caption{Illustration of shape-wise structured pruning. The F kernels are split into multiple groups, kernels in the same group are pruned on the same locations.}\label{pruning}
\end{figure}

In conclusion, we make the following contributions: 1) we propose a sparse wise dataflow to skip the cycles of processing Multiply-and-Accumulates (MACs) that have zero weights and minimize energy through zero gating to further avoid unnecessary computations. 2) we propose a Vector Generator Module to reuse and generate the necessary input activations for sparse CNNs. 3) we co-design both the accelerator architecture and the loop tiling to minimize off-chip memory accesses and maximize performance by slicing the input feature maps to best match the capacity of Block Random Access Memory (BRAMs) on FPGA.

Experiments demonstrate that the proposed accelerator can achieve 987\,imag/s and 48\,imag/s performance for AlexNet and VGG-16 on Xilinx ZCU102, respectively, which provides 1.5$\times$ to 6.7$\times$ speedup and 2.0$\times$ to 6.2$\times$ energy-efficiency over previous CNN FPGA accelerators

\section{Background}
CNNs consist of multiple types of layers, including  convolutional layers, pooling layers and fully-connected layers. Through these layers, inputs are processed and propagated, thus to be classified or recognized. The convolution operation uses an $R\times R$ window to slide through the input feature map to extract features. At every location, the input activations inside the window are multiplied by corresponding weights and the products are accumulated to compute the partial sum of an output activation. Note that the partial sums in different input channels are accumulated to compute the output activation.

\begin{table}
	\caption{SPARSITY AND ACCURACY COMPARISON}
	\label{tab:Pruningcomparison}
	\begin{tabular}{lllll}
		\toprule
		\multirow{2}{*}{model} & \multicolumn{2}{c}{Deep Cmp} & \multicolumn{2}{c}{Coarse-grained pruning} \\
		& Sparsity (\%) & Top1-E (\%) & Sparsity (\%) & Top1-E (\%) \\
		\midrule
		AlexNet & 11.15 & 42.78 & 11.03 & 42.72 \\
		VGG-16 & 7.61 & 31.17 & 8.07 & 31.33 \\
		\bottomrule
	\end{tabular}
\end{table}

\subsection{Network Pruning}
CNNs have achieved remarkable success in various applications~\cite{hejia_isrr19,he2015delving,girshick2014rich,redmon2016you,zhang2019robot} at the cost of huge amount of computations. The weights pruning method has been proved as an effective solution to reduce the computation and memory burden of these models without significant  accuracy loss~\cite{han2015learning,han2015deep,ding2017c, deng2018permdnn, zhou2018cambricon, zhang2018adam}. Compared to unstructured pruning techniques, structured pruning techniques are less irregular and more hardware friendly at the cost of a slightly lower compression rate. With shape-wise structured pruning (cf. Fig.~\ref{pruning}), kernels in the same group will be encoded in the same compressed format. By sharing a common compressed format, the irregularity of sparse weights will be reduced. Besides, the load of each parallel processing engines is balanced. Table \ref{tab:Pruningcomparison} shows the sparsity and accuracy comparison of Deep Compression~\cite{han2015deep} and Coarse-grained pruning~\cite{zhou2018cambricon}. The difference of sparsities between Deep Compression and  Coarse-grained pruning is almost negligible. Although structured pruning techniques reduce the irregularity of sparse weights, previous FPGA CNN accelerators still can not exploit weight sparsity efficiently. To address the remaining irregularity, we propose a sparse wise dataflow and propose a set of architecture optimization techniques.


\begin{figure}
	\centering
	\includegraphics[scale=0.38]{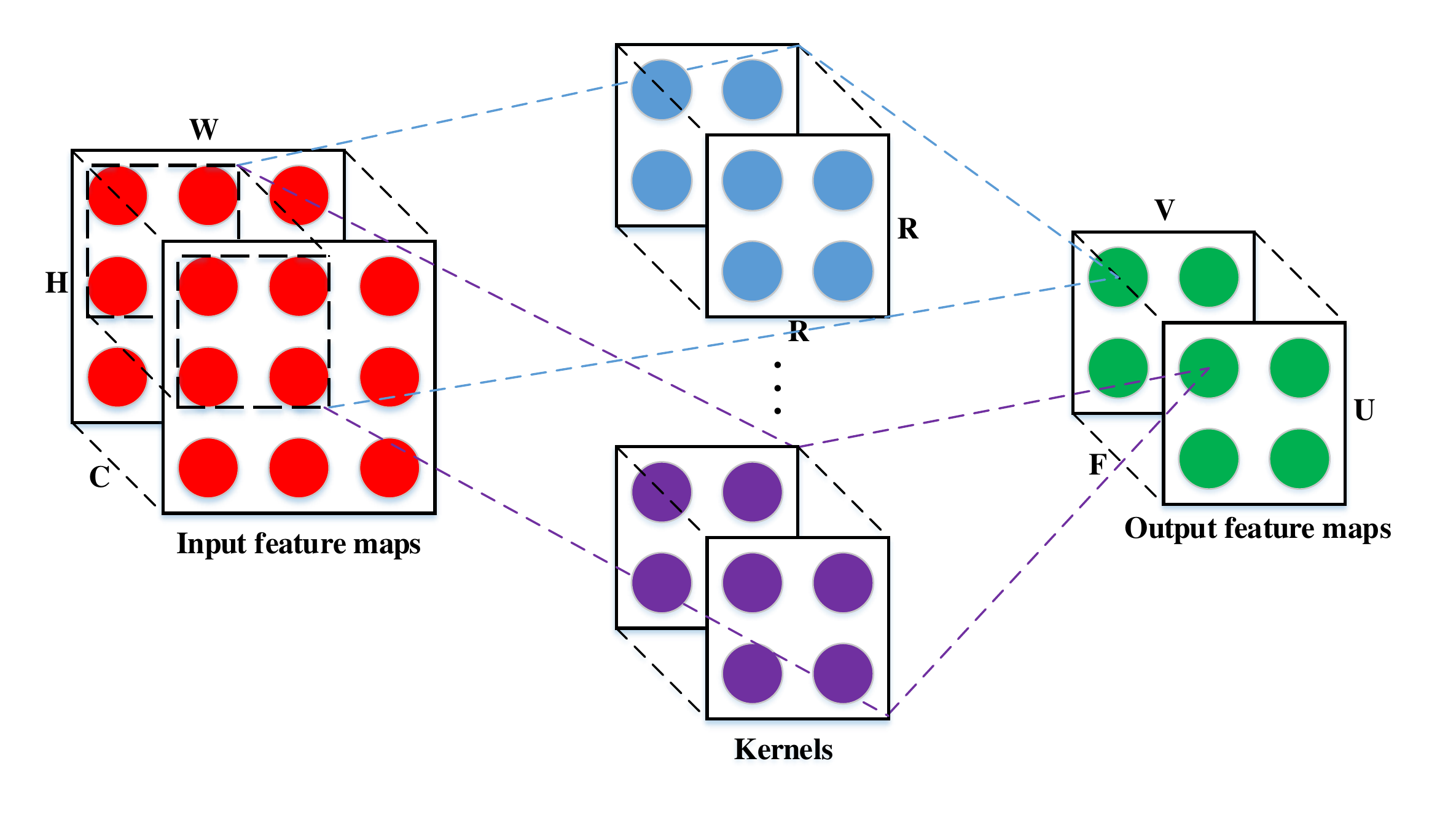}\\
	\caption{The logical graph of convolutional operation. Convolutional layers conduct 2D convolutions on a set of input feature maps and add the results to get output feature maps.}\label{conv_operation}
\end{figure}

\begin{algorithm}
	\scriptsize
	\SetAlgoNoLine 
	\SetInd{1em}{0.5em}
	\SetKwInOut{Input}{\textbf{Input}}\SetKwInOut{Output}{\textbf{Output}} 
	\BlankLine
	\For{f = 0; $f \textless F$; f++}{
		\For{c = 0; $c \textless C$; c++}{
			\For{u = 0; $u \textless U$; u++}{
				\For{v = 0; $v \textless V$; v++}{
					\For{kh = 0; $kh \textless R$; kh++}{
						\For{kw = 0; $kw \textless R$; kw++}{
							$Y^f_{u,v} += W^{f,c}_{kh,kw} \times X^c_{u+kh,v+kw}$; 	
						}
					}
				}
			}
		}
	}
	\caption{Pseudo code of convolutional layer}
	\label{Algorithm1}
\end{algorithm}

\subsection{Loop Operation}
In a typical CNN, convolutional layers take up about 90\% of the computation in inference procedure. A convolutional layer can be characterized by six parameters: $F$, the number of output feature maps (output channels); $C$, the number of input features maps (input channels); $U$, the height of output feature map; $V$, the width of output feature map; $R$, the kernel size; $S$, the stride size, as shown in Fig.~\ref{conv_operation}. The computation can be described in a deep nested loop as illustrated in Algorithm \ref{Algorithm1}. Feature map related loops $f$ and $c$ index the output and input channels, respectively. Activation related loops $u$ and $v$ index each activation of feature maps. Finally, weight related loops $kh$ and $kw$ index weights of each kernel. Obviously, convolution operation exhibits high parallelism in channel, activation and weight levels.

To achieve high performance, the above deep nested loop is unrolled and mapped to a parallel hardware. This involves loop tiling and loop interchange strategy\cite{guo2017survey}. Loop tiling keeps all the input data of a loop tile stored on chip to reduce external memory access. External memory access happens when the operation of a new tile begins. Besides, a part of data may be temporally reused across the adjacent tiles. Loop interchange strategy decides the order of the loop tiles. Both loop tiling and loop interchange strategy decide the dataflow of hardware.

\section{Sparse wise Dataflow}
There have been dense CNNs dataflows on FPGA\cite{zhang2017improving, li2017fpga, ko2017design}. However, these dataflows cannot leverage the benefits of sparse CNNs since most multiplication operations involve zero weights that will not contribute to the corresponding output feature map. Therefore, it is extremely essential to skip the cycles of processing MACs that have zero weights.

Recently, designing dataflows for sparse CNNs on ASIC platforms has attracted attentions from the research community. However, these dataflows will not be efficient for FPGA platforms due to the architecture difference between ASIC and FPGA. For example, SCNN architecture \cite{parashar2017scnn} applies input-stationary dataflow where the inner computation is a Cartesian production. In SCNN architecture, there are $N$ PEs and each PE contains an $I\times F$ multiplier array. Consequently, it simultaneously requires $N\times I$ input activations and $N\times F$ weights for computation. Although, this dataflow temporally reuses input activations, it still requires to update $N\times F$ weights in each cycle. Furthermore, this method firstly requires significant coordinates computation to locate output activations. Then using Cartesian production, this dataflow returns multiple partial sums, which are needed to be arbitrated before being saved. Cambricon-S\cite{zhou2018cambricon} applies output-stationary dataflow where the inner computation is a Vector dot product. Cambricon-S implements a centralized indexing module to select input activations and only transfers selected activations and indexes to PEs without extra coordinates computation. However, this dataflow only performs parallel computation in channel dimensions by spatially sharing input activations, thus requires $M$ input activations and $N\times M$ weights for computation, which will lead to poor parallelism on FPGAs.

We propose a sparse wise dataflow to accelerate CNNs on FPGA (cf. Alg.~\ref{Algorithm2}). For a convolutional layer with input feature maps Ifamp$(C,H,W)$ and kernels Kernel$(F,C,R,R)$,  we select and fetch a vector of $T_{oc}$ nonzero weights from $T_{oc}$ adjacent kernels at each access. Similarly, we also fetch a vector of $T_{om}$ activations which are in the same row of input feature map. Meanwhile, we compute $T_{oc}$ element-vector mutiplications on the PE array. After shape-wise structured pruning, the coordinates of nonzero weights in different kernels are uniformed. So each element-vector mutiplication involves a weight and a shared vector of $T_{om}$ activations. For example, in the step 1, the weight vector is $[W^{0,0}_{0,0}, \cdots,W^{T_{oc},0}_{0,0}]$, and the input vector is $[X^0_{0,0}, \cdots,X^0_{0,T_{om}}]$. In the step 2, we fetch weight vector $[W^{0,0}_{0,1}, \cdots,W^{T_{oc},0}_{0,1}]$ and input vector $[X^0_{0,jump}, \cdots,X^0_{0,T_{om}+jump}]$ for next computation and compute $T_{oc}$ element-vector mutiplications in pipeline mode. The variable jump equals to the number of zeros between adjacent nonzero weights in the same row of a kernel.

\begin{figure}
	\centering
	\includegraphics[scale=0.6]{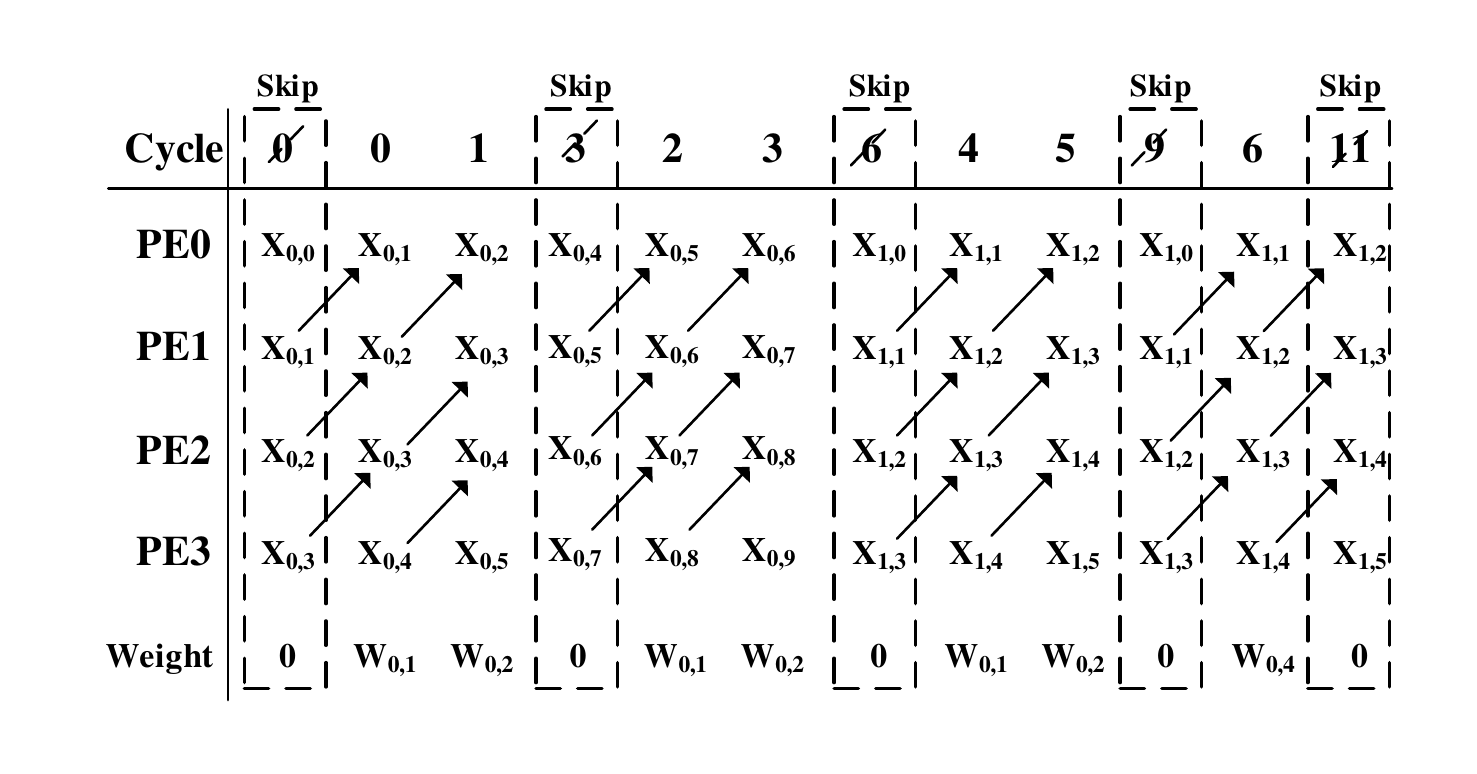}\\
	\caption{Sparse wise dataflow for CONV layers. When weights equal to zero, the cycles of processing MACs will be skipped by controlling the upper bound of corresponding loop illustrated in Algorithm \ref{Algorithm2}.}\label{dataflow_conv}
\end{figure}

Convolution windows that produce adjacent outputs share part of input activations. Therefore, the two successive fetched input vectors from the same row of input feature map involve partially shared data. We add a dedicated Vector Generator Module (will be introduced in the next section) to update the input register for reusing data. We describe the convolution operation execution pattern in Fig.~\ref{dataflow_conv}. Using the pattern in Fig. \ref{dataflow_conv}, the four PEs compute four adjacent output activations and skip the cycles of processing MACs that have zero weights.

\IncMargin{0.5em} 
\begin{algorithm}
	\scriptsize
	\SetAlgoNoLine 
	\SetInd{1em}{0.5em}
	\SetKwInOut{Input}{\textbf{Input}}\SetKwInOut{Output}{\textbf{Output}} 
	
	\Input{
		\\
		The nonzero weight array $W^{oc,ic}_{kh,kw}$\;\\
		The index array $Index[\ ]$\;\\
		The row pointer of index $R\_pointer[\ ]$\;\\
		The number of nonzero weights in each input channel $Offset[\ ]$\;\\
		The input activation $X^{ic}_{ih,iw}$\;
		The Kernel size R$\times$R$\times$C$\times$F\;\\
		The Output feature map size U$\times$V$\times$F\;}
	\Output{
		\\
		The output activation $Y^{oc}_{oh,ow}$\;}
	\BlankLine
	
	Initialize the  parameter $kw = 0$ , $kh = 0$, $i=0$ and $jump = 0$\; 
	\For{oh = 0; $oh \textless U; oh = oh + U_{t}$}{
		\For{oc = 0; $oc \textless F$; oc = oc + N}{
			\For{ic = 0; ic \textless C; ic++}{
				\If{$ic \textgreater 0$}{
					$i = i + Offset[ic-1]$\;}
				\For{$T_{oh} = oh; T_{oh} \textless oh+U_t; T_{oh}++$}{
					\For{ow = 0; ow \textless V; ow = ow + M}{
						\If{$kh = R \&\&  ow\textgreater 0$}{
							$i = i - Offset[ic]$\;}
						\For{kh = 0; kh \textless R; kh++}{
							$kh\_count = ic \times R + kh$\;
							$kw\_max = R\_pointer[kh\_count]$\;
							\If{$kh \textgreater 0$}{
								$i = i + R\_pointer[kh\_count-1]$\;}
							$jump = 0$\;
							\For{kw = 0; kw \textless kw\_max; kw++}{
								$jump = jump + Index[i + kw]$\;
								//Loop unroll\\
								\For{$T_{oc}=oc;T_{oc} < oc+N;T_{oc}++$}{
									\For{ $T_{ow} = ow; T_{ow} \textless ow + M; T_{ow}++\quad$}{
										$Y^{T_{oc}}_{T_{oh},T_{ow}} = W^{T_{oc},ic}_{kh,kw} \times\ X^{ic}_{T_{oh}+kh,T_{ow}+kw+jump}$\;}	
								}
							}
						}
					}
				}
			}
		}
	}
	\caption{Pseudo code of our dataflow}
	\label{Algorithm2}
\end{algorithm}

\begin{figure}
	\centering
	\includegraphics[scale=0.65]{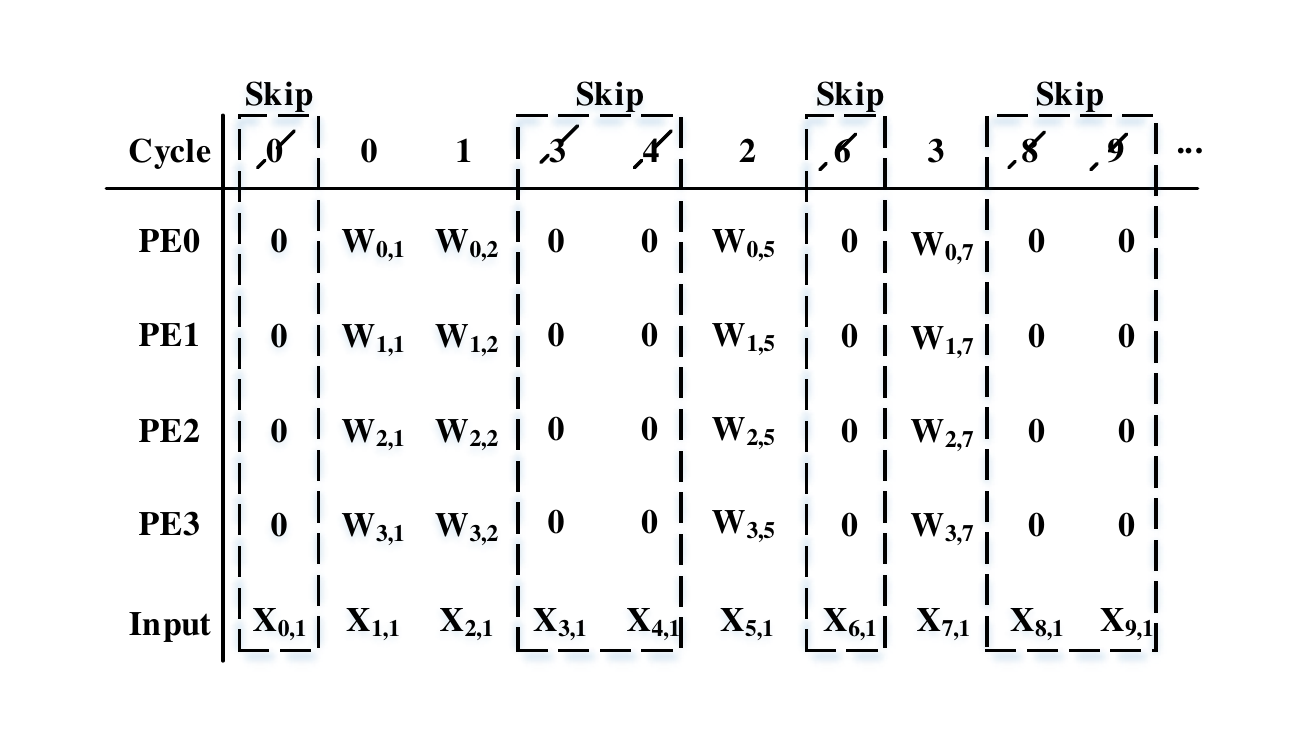}\\
	\caption{Execution pattern of FC layers. Similar to CONV layers, the cycles of processing MACs will be skipped when weights equal to zero.}\label{dataflow_fc}
\end{figure}

For a fully-connected layer with input vector $I\_vector(C,1)$ and weight matrix $W(F,C)$,  the weight buffer delivers a vector of $T_{om}$ nonzero weights from the adjacent rows of matrix $W$ at each access (see Fig.~\ref{dataflow_fc}). The input register only delivers a single activation.



\begin{figure}
	\centering
	\includegraphics[scale=1.2]{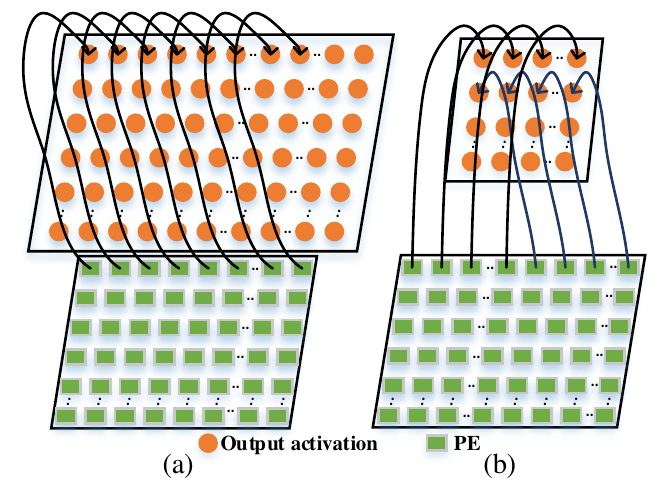}\\
	\caption{The two situations of network mapping. (a)When V$\geq $M, one Processing Unit (PU) computes M partial sums of M output activations which are in the same row; when V$\textless$M, one PU computes M partial sums of M output activations which are in the different rows.}\label{mapping}
\end{figure}

This dataflow requires a group of buffers to save partial sums produced by convolution operations. Considering a convolutional layer with output feature map Ofamp$(F,U,V)$, we applied a $N\times M$ PE array. Make $T_{oc} = N$ and $T_{om}=M$, to prevent partial sums from being saved to and restored from DRAM, the depth of the partial sum (Psum) buffer should be at least $\lceil U\times V /M\rceil$. However, some CNNs like VGG-16 involve a quite large product of $U\times V$ in the first two layers. Therefore, we partition the input feature map into $\lceil H/H_t \rceil$ tiles, where  
\begin{equation}\label{tile}
\begin{aligned}
H_t &= (U_t - 1)\times Stride + R \\
&= (\lfloor \frac{M\times Slice_{BRAM}}{V}\rfloor - 1)\times Stride + R
\end{aligned}
\end{equation}
The parameter $Slice_{BRAM}$ represents the capacity of one block of BRAM on FPGA. The adjacent tiles share ($R-Stride$) rows because the sliding-window nature of the convolution operation introduces data dependency at tile edges.

When the width of output feature map $V \textless M$, we map a $\lfloor M/V \rfloor \times V$ tile to one PU for efficiency because each PE computes one output activation. Otherwise, we map a $1\times M$ tile to one PU, as shown in Fig.~\ref{mapping}. The PE arracy consists of N PUs and each PU consists of M PEs in the same row. The input activations are shared across PUs. Different PUs compute output activations which are from different output channels.

By spatially sharing both input activations and weights, and, temporally reusing partial input activations, we reduce the bandwidth requirement to less than $N\times B+M\times A$. Therefore, the bandwidth requirement of the accelerator is much smaller than Cambricon-S ($M\times A + N\times M \times B$) and SCNN ($N\times F \times B$), where $A$ and $B$ represent the widths of activation and weight, respectively.

\section{Architecture of Sparse Wise Accelerator}
\label{sec:guidelines}
In this section, we introduce the detailed architecture of our accelerator, to address the remaining irregularity of shape-wise pruned CNNs.

\begin{figure}
	\centering
	\includegraphics[scale=0.35]{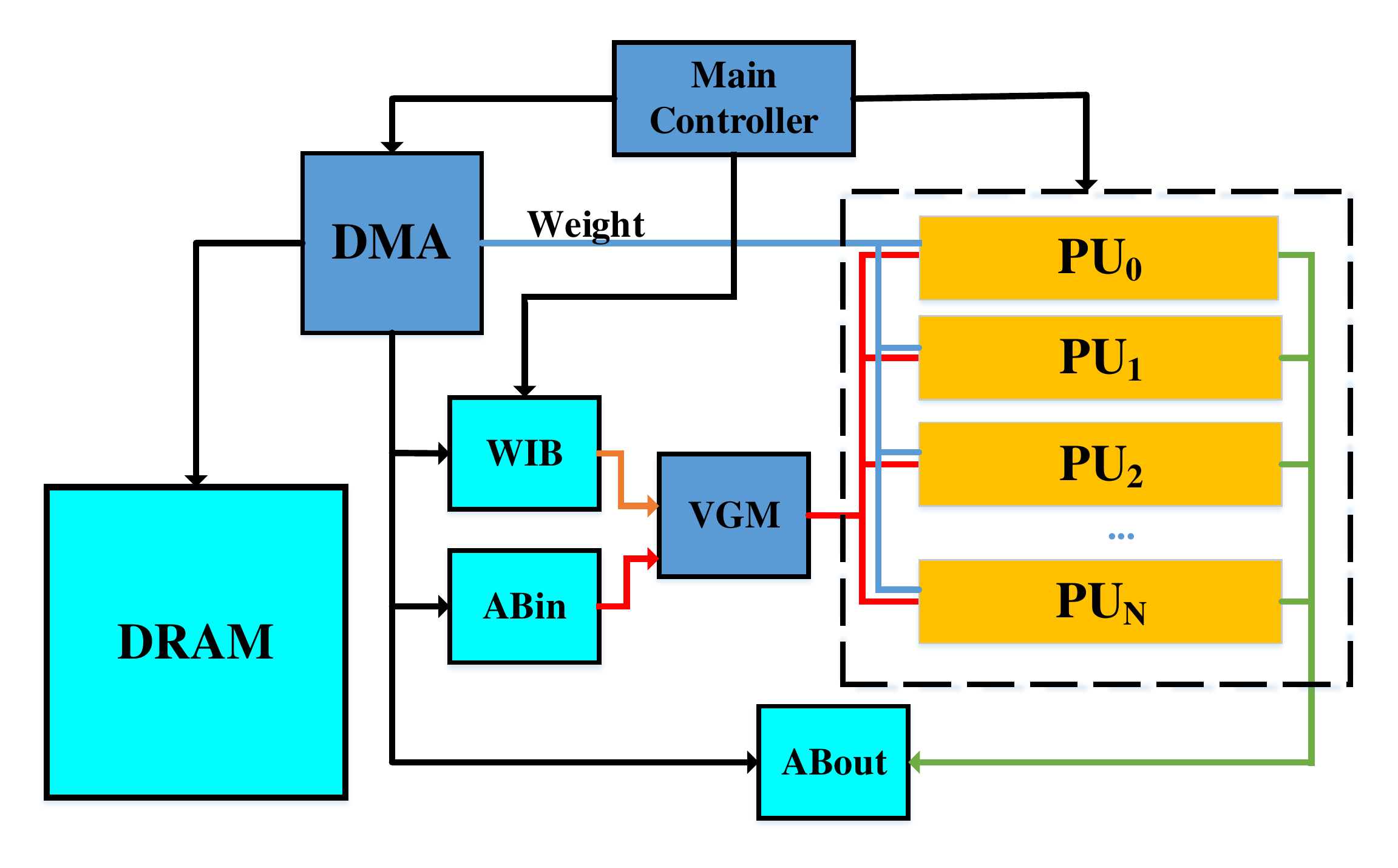}\\
	\caption{Accelerator Architecture. The DRAM is implemented with Double Data Rate Random Access Memory in Processing System on Xilinx FPGA.}\label{archicture}
\end{figure}

{\bfseries Overview.} Fig. \ref{archicture} depicts the overall block diagram of our proposed accelerator. Following the proposed dataflow, we design a Vector Generator Module (VGM) to address the sparsity with shared indexes. We design a PU  that has multiple PEs to compute adjacent output activations in parallel. Multiple PUs constitute an array to compute multiple output activations across output channel in parallel. The storage module consists of two activation buffers (ABin and ABout), a Weight Index Buffer (WIB), $N$ Weight Buffers (WBs), and $N\times M$ Partial Sum Buffers (PSBs). The Main Controller decodes instructions and weight indexes into detailed control signals for all other modules.


\subsection{Addressing with sparsity}
The accelerator is designed to exploit structured sparsity for performance gain and energy reduction. In our accelerator, sparsity is processed by VGM and PU together. The VGM receives input activations from ABin and decodes weight indexes from WIB, then produces the selected activations that are broadcast to all the PUs. Meanwhile, these input activations will be cached for next selection in VGM since there is overlapping when the kernel slides across the input feature map. Each PU receives the read address of weight from Main Controller and reads out the needed weight following the principles described in proposed dataflow, thus avoiding unnecessary computations.




\begin{figure}
	\centering
	\includegraphics[scale=0.5]{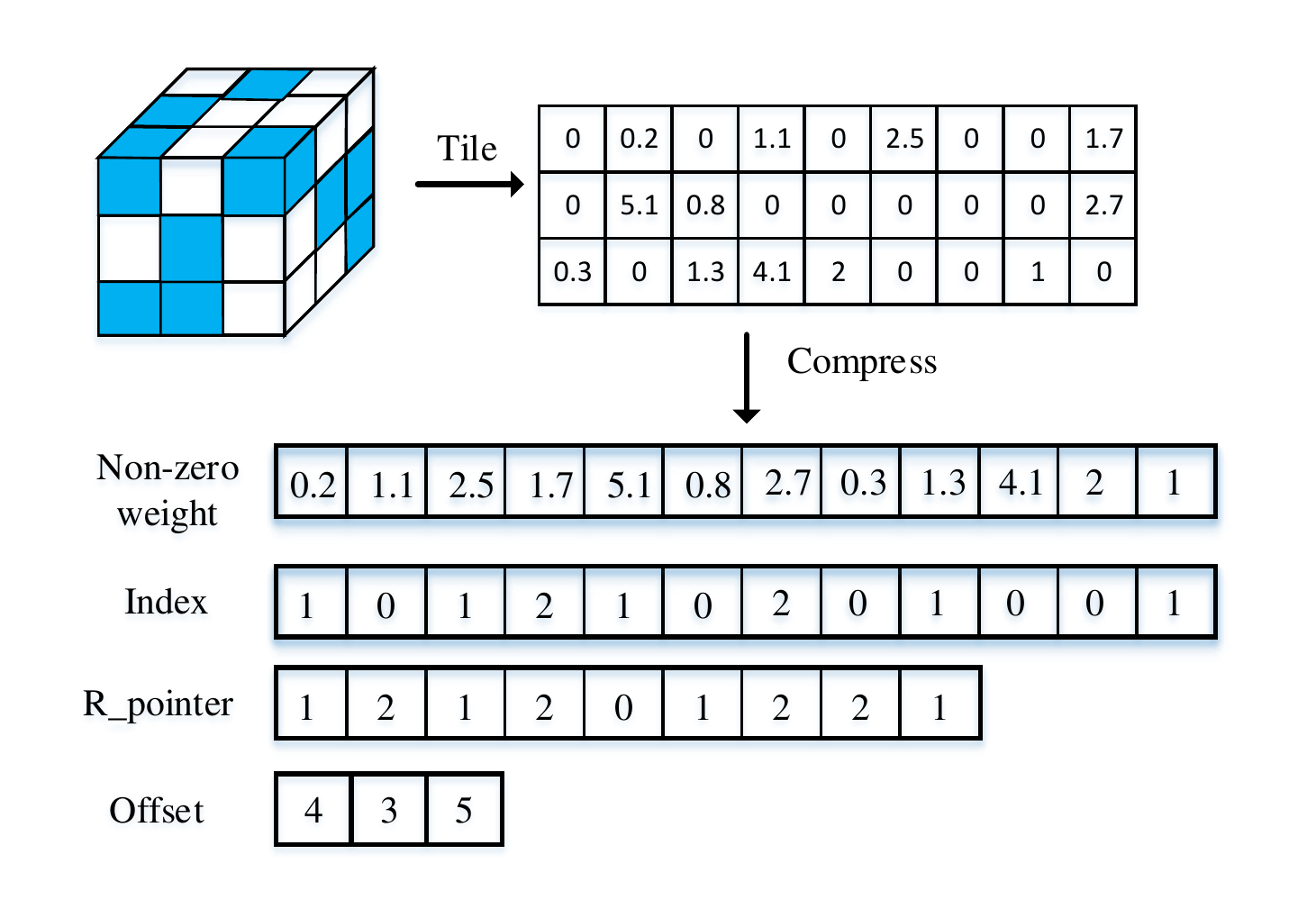}\\
	\caption{Index representation of weights in CONV layers. The Index represents the number of pruned weights between two nonzero weights. The r\_pointer represents the number of remaining weights in each row. The Offset shows the remaining weights in each channel of one kernel.}\label{conv_index}
\end{figure}

\begin{figure}
	\centering
	\includegraphics[scale=0.45]{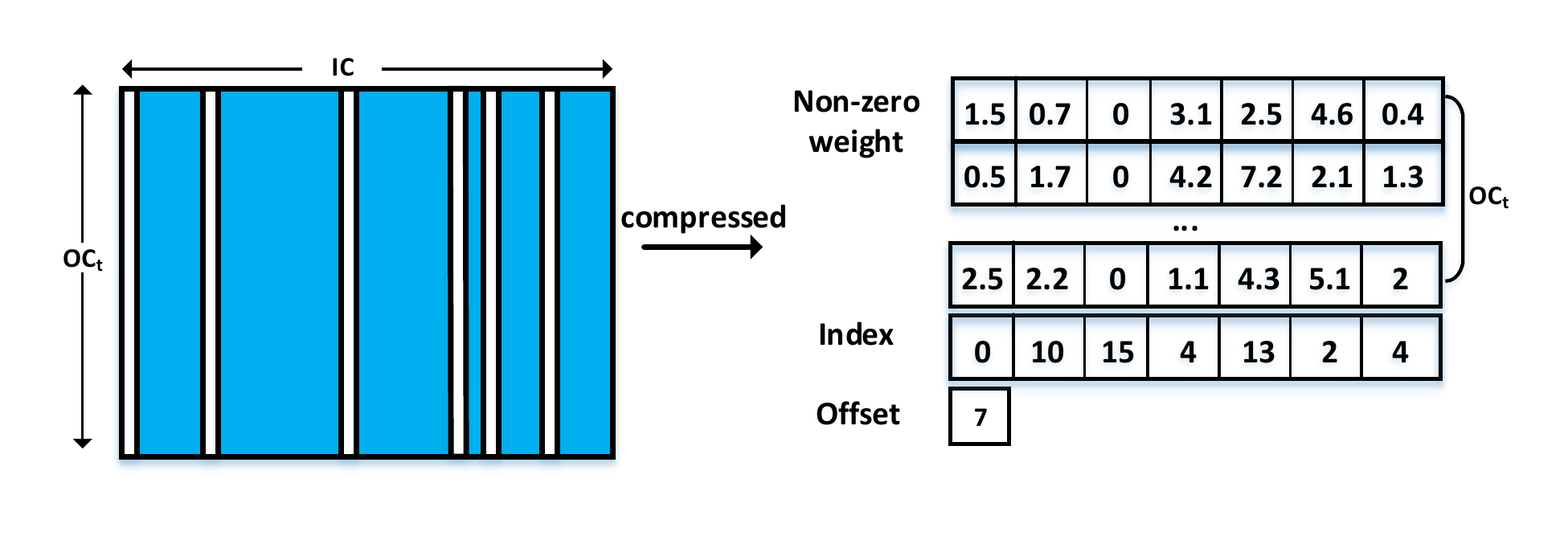}\\
	\caption{Index representation of weights in FC layers. Padding filler zero to prevent overflow.}\label{fc_index}
\end{figure}

\begin{figure*}
	\centering
	\includegraphics[scale=0.55]{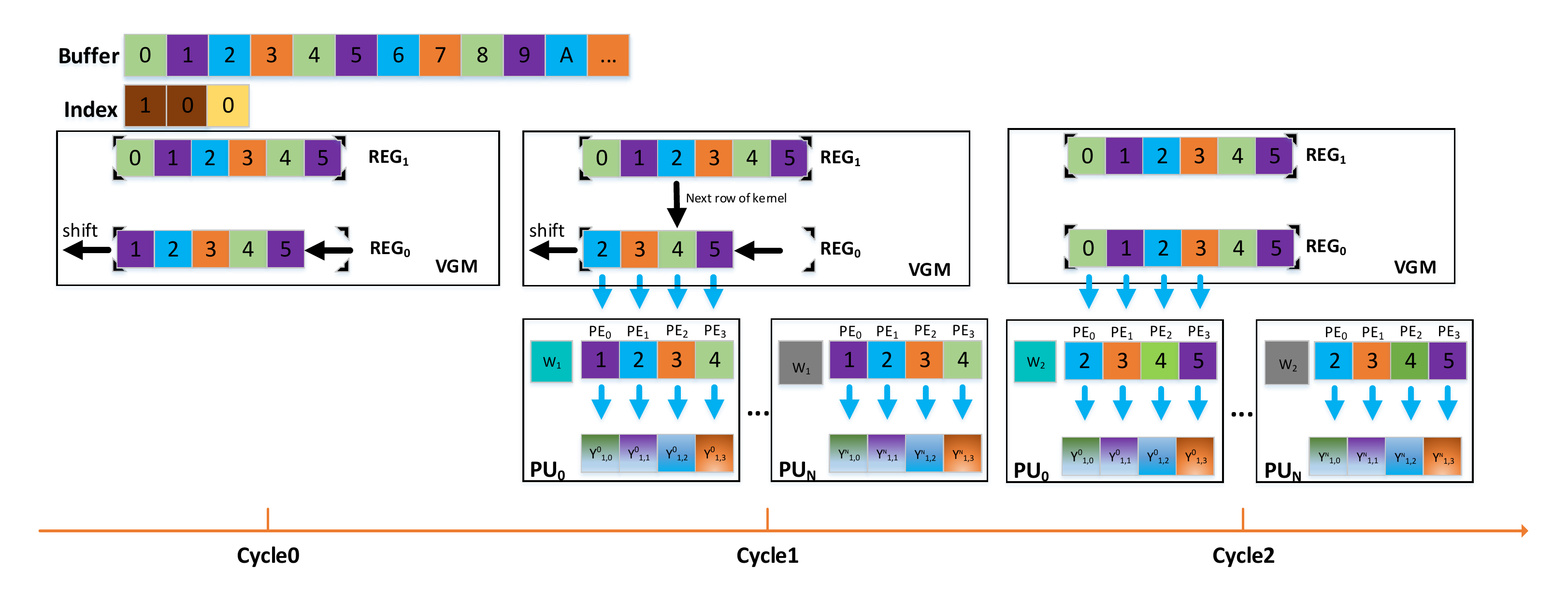}\\
	\caption{Vector Generator Module. The VGM buffers and selects input activations to reuse them and to address the weight sparsity. It is shared by all the PUs.}\label{VGM}
\end{figure*}

{\bfseries Index.} Before elaborating the VGM and PU, we clearly explain how we store and index the sparse weights. We store the sparse weights that result from shape-wise pruning using Compressed Sparse Row (CSR) format, which only requires $2a+R\times C+C$ numbers, where $a$ is the number of nonzero weights, $R$ is the number of rows and $C$ is the number of input channels. Owing to shape-wise pruning, each kernel shares the common index. To compress further, we store the step index instead of the absolute position. We encode both step index and $R\_pointer$ in 4 bits, and encoded $Offset$ in 16 bits as illustrated in Fig.~\ref{conv_index}.

The 4 bits index is large enough in convolutional layers, however, the situation is different in fully-connected layers. When we need an index larger than the bound, we will pad a filler zero to prevent overflow. Regarding the example in Fig.~\ref{fc_index}, when the step index exceeds the largest 4-bit unsigned number, we pad a filler zero.

{\bfseries VGM.} The VGM module processes the sparsity by selecting the needed input activations and transfers the selected input activations to all the PUs, see Fig. \ref{VGM}. We design a central VGM shared by multiple PUs to more efficiently process shared indexes from structured sparsity. For example, firstly, when index is "1", the activations in register $REG_0$ will be shifted by $A$ bits to the left (the most left A bits, where A represents the width of activation), and the data with index $(REG_0[0],REG_0[Stride],...REG_0[(M-1)\times Stride])$ will be broadcasted to all the PUs. In $cycle1$,  when index is "0", the activations in register $REG_0$ will be further shifted by $A$ bits. PUs do MACs with previous selected input activations and cache the activations selected by VGM in this cycle. In $cycle2$, the row number of kernel---$kh$ (see Algorithm \ref{Algorithm2}) increases. Therefore, $REG_0$ reloads activations from $REG_1$ and does data-shift according to the new weight index. The depths of both $REG_0$ and $REG_1$ are set to $(M-1)\times stride+R$. To leverage the overlap between activation selection and memory access, the activations in $REG_1$ will be updated after $REG_0$ lastly reloading activations from $REG_1$.

As PUs share the same indexes of weights due to shape-wise pruning, the module for selecting activations (VGM) is shared by all the PUs, thus reducing the indexing module overhead and bandwidth requirement between VGM and PUs.



\begin{figure}
	\centering
	\includegraphics[scale=0.45]{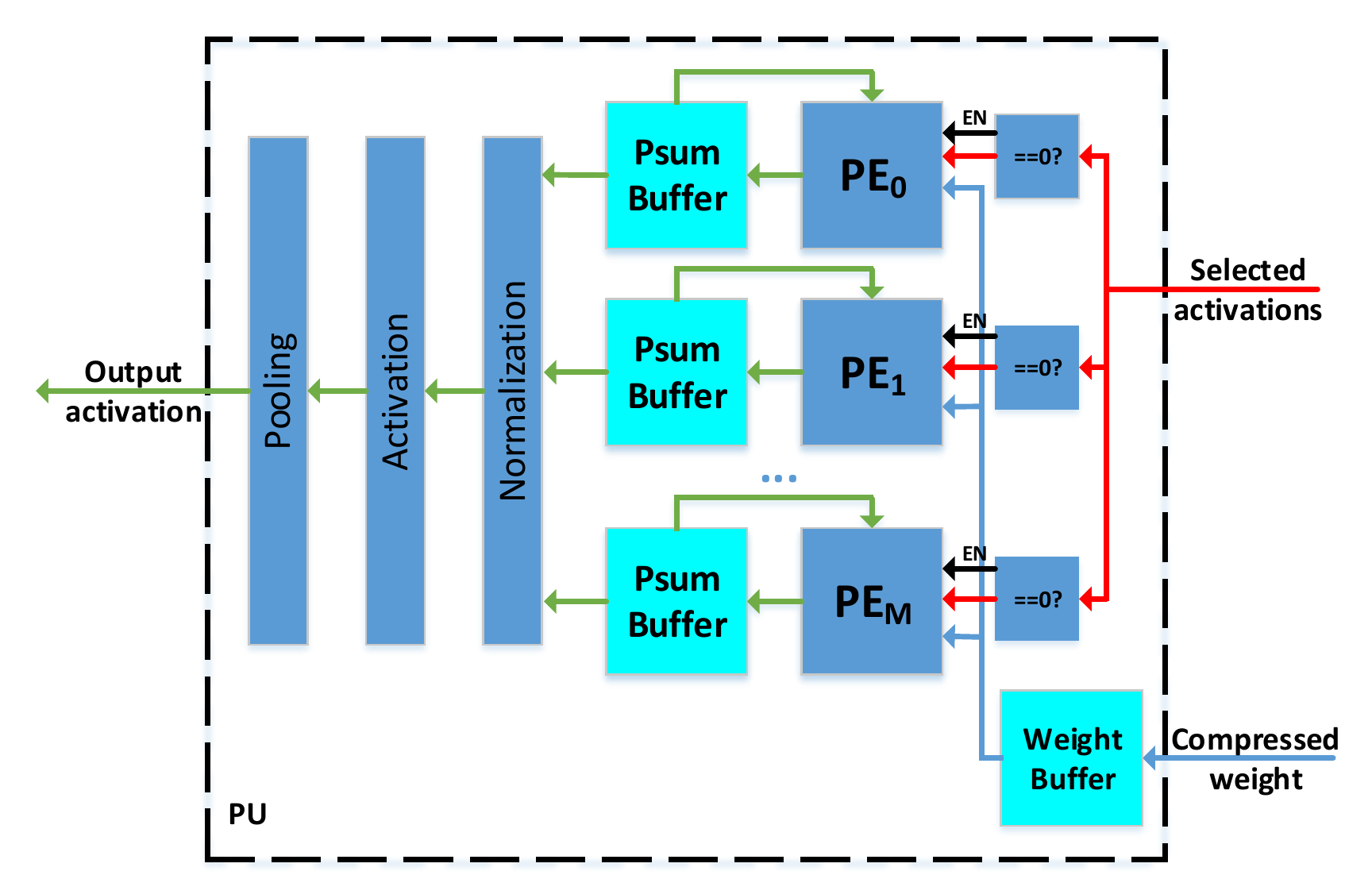}\\
	\caption{The architecture of the PU. The PU processes all operations in CNNs. It contains M homogeneous PEs, and the number of PEs can be configured according to the CNN model and FPGA platform.}\label{PU}
\end{figure}

{\bfseries PU.} The PU processes all operations in CNNs. Each PU consists of M zero-value discriminators, M homogeneous PEs, M homogeneous PSBs, a WB, pooling, normalization, and activation module, see Fig.~\ref{PU}. Weights can be stored separately in PUs as output activations from different channels involve independent kernels.  A selected activation firstly streams into a zero-value discriminator, and then will be used in PE if it does not equal to zero. Meanwhile, the discriminator produces an enable signal to control the clock gating of PE. When the selected activation equals to zero, the corresponding PE is clock gated. To minimize data communication, the partial sums produced by PE will be saved in a local partial sum buffer which is placed next to the PE. Until the entire computation of an output activation is done, the final partial sum will be transfered to activation and pooling module. 

As there is neither spatially sharing nor temporally reuse for weights in fully-connected layer, it requires quite large input bandwidth. Although all the PUs can be active in fully-connected layer, the required off-chip memory bandwidth  cannot be satisfied on the FPGA platform. Thus, we only keep one PU active when the M is large enough in fully-connected layer.



\subsection{Optimize PE for quantization}
One of the most commonly methods for model compression is to quantize both weights and activations. The low-bit activations and weights require small bandwidth, which benefits to improve the throughput if the accelerator is a computation bounded design\cite{zhang2015optimizing}. However, the proposed design is BRAM limited. Directly increasing the size of PE array will lead to a failure of implementation on FPGA because the number of required BRAM will easily exceed the available number. Therefore, we maintain the datapath of activations and optimize the structure of PE. For example, the activations and weights are quantized to 8-bit. We dispatch two selected 8-bit activations into one PE. Due to the proposed dataflow, the two activations multiply with the same weight in two DSP slice, respectively. Then the partial sums can be concatenated and stored in the buffer.

\subsection{Storage}
As the data processed in our accelerator have different behaviors, we split storage into five parts: an ABin, an ABout, a WIB, $N$ WBs, and $N\times M$ PSBs.

For the ABin and ABout, we set the width as $16\times (M-1+R)$-bit and $16\times M$-bit, respectively, so as to provide $(M-1+R)$ input activations for VGM and to fetch $M$ output activations from PU at each access. Benefit from the proposed dataflow, $N$ PUs share the same input activations, we read input activations and produce output activations row by row. Thus, we set a small depth for both ABin and ABout.

For the WB in each PU, we use a dual port ram that we select the read width as $16$-bit for one port and $16\times M$-bit for the other port. In particular, the write width of both two ports is $16\times M$-bit. In the convolutional layer, only one weight will be read out and be broadcasted to all the PEs. Whereas in the fully-connected layer, $M$ weights will be fetched and be transfered to the corresponding PE.

For the WIB, we select the width as $16$-bit as we use CSR format where we deploy $16$-bit, $4$-bit, $4$-bit for $Offset$, $Index$, and $R\_pointer$, respectively. Thus, the $Index$ and $R\_pointer$ are stored aligning to $4$ bits. We divide the WIB into three parts for the three components of the compressed weights index.

For the PSB, we set the width to $32$-bit. We store the partial sums and cache output activations in PSB, as the ABout fetchs output activations from PUs in turns. We map each PSB to one BRAM. Because if we map each PSB to two or more BRAMs, the on-chip memory resources will easily be the bottleneck of the available peak performance, leading to the damping of throughput.

The size of the five buffers are decisive to overall performance and energy consumption. For example, the size of PSB decides the number of tiles. Small size of PSB requires large number of tiles that leads to costly off-chip memory accessed. Whereas large size of PSB leads to unscalability for small layers in CNNs. Thus, we generally deploy 2KB, 2KB, 4KB, 512B and 2KB for ABin, ABout, WIB, WB and PSB, respectively. The configuration of these buffers can be adjusted accroding to the CNNs.

\section{Experiment}

\subsection{Experiments Setup}
\setlength\parindent{1em} We evaluate our design on the Xilinx ZCU102 evaluation kit consisting of an Ultrascale FPGA, quad ARM Cortex-A53 processors, 4GB PS DDR4 and 512MB PL DDR4. In this work, we use verilog for RTL implementation and employ Xilinx Vivado (v2017.2) to compile the source code to bitstream. The design method is inspired by the DNNWEAVER\cite{dnnweaver2016}. Our FPGA implementation is synthesized at 200MHz frequency. We use a GPIO to USB adapter to read the power directly from the PMbus in the FPGA board. We comprehensively apply~\cite{zhou2018cambricon, zhang2018adam} methods to train the CNN models. In Our experiment, we test typical CNNs including Lenet, Alexnet and VGG-16 and achieve $11.85\%, 32.92\%, 36.75\%$ sparsity of Lenet, Alexnet and VGG-16 without significant accuracy loss.

\begin{table}[h]
	\caption{RESOURCE UTILIZATION BREAKDOWN}
	\label{tab:breakdown}
	\scalebox{1.0}{
		\begin{tabular}{|c|c|c|c|c|}
			\hline
			& BRAM & LUT & DSP & FF\\
			\hline
			PU & 28 & 6407 & 28 & 5463 \\
			\hline
			DMA & 111 & 26875 &  0 & 6940 \\
			\hline
			Controller & 0 & 28392 & 6 & 1282 \\
			\hline
			VGM & 0 & 27963 & 0 & 7285 \\
			\hline
			ABin & 2 & 0 & 0 & 0 \\
			\hline
			ABout & 1 & 0 & 0 & 0 \\
			\hline
			WIB & 2 & 0 & 0 & 0\\
			\hline
			Total & 1460(80\%) & 390K(65\%) & 1350(53\%) & 278K(51\%) \\
			\hline
			Available & 1824 & 600K & 2520 & 550K \\
			\hline
		\end{tabular}
	}
\end{table}

\newcolumntype{I}{!{\vrule width 2pt}}
\newlength\savedwidth
\begin{table*}[t!]

	\caption{PERFORMANCE COMPARISON WITH PREVIOUS IMPLEMENTATION}
	\label{tab:freq}
	\scalebox{0.9}{
		\begin{tabular}{|c|c|c|cIc|c|cIc|c|}
			\hline
			&\cite{zeng2018framework}&\cite{kala2019high}&\cite{lu2019efficient}&Ours&\cite{ma2017optimizing}&\cite{lu2019efficient}&Ours&Ours\\
			\hline
			CNN type &  Alexnet & Alexnet & Alexnet & Alexnet & VGG-16 & VGG-16  & VGG-16 & VGG-16\\
			\hline
			Device & Zynq ZC706 &  XC7VX690T & Zynq ZCU102 & Zynq ZCU102 & Zynq ZCU102 & Arria-10 GX1150  & Zynq ZCU102 & Zynq ZCU102\\
            \hline
            Accelerator type & sparse & dense & sparse & sparse & sparse & dense  & sparse & sparse\\
			\hline
			Frequency(MHz) &  - & 200 & 200 & 200 & 200 & 150  & 200 &200\\
			\hline
			Precision & - & 16bit fixed & 16bit fixed & 16bit fixed & 16bit fixed & 16bit fixed  & 16bit fixed & 8bit int\\
			\hline
			DSP Utilization &  - & 1436(40\%) & 1144(45\%) & 1350(53\%) & 1144(45\%) & 1518(100\%)  & 1350(53\%) & 2520(100\%)\\
			\hline
			Logic Utilization &  - & 468K(67\%) & 552K(92\%) & 390K(65\%) & 552K(92\%) & 161K(38\%)  & 390K(65\%) & 405K(67\%) \\
			\hline
			BRAM &  - & 423(39\%) & 912(48\%) & 1460(80\%) & 912(48\%) & 1900(70\%)  & 1460(80\%) & 1460(80\%)\\
			\hline
			Performance(imag/s) &  147 & 548 & 640 & 987 &  21 & 22  & 48 & 96\\
			\hline
			Power(W) &  9.6 & 17.3 & 23.6 & 15.4 & 23.6 & 45.0  & 15.4 & 17.1\\
			\hline
			Efficiency(imag/s/W) &  15.37 & 31.71 & 27.13 & 64.13 & 0.92 & 0.50  & 3.11 & 5.61\\
			\hline
		\end{tabular}
	}
\end{table*}

\begin{table*}[t!]

	\caption{PERFORMANCE DENSITY COMPARISON WITH PREVIOUS IMPLEMENTATION}
	\label{tab:per_eff}
	\scalebox{0.9}{
		\begin{tabular}{|c|cIc|c|c|cIc|c|}
			\hline
			&\cite{kala2019high}&Ours&\cite{lu2019efficient}&\cite{lian2019high}&\cite{ma2017optimizing}&Ours&Ours\\
			\hline
			CNN type & Alexnet & Alexnet & VGG-16 & VGG-16 & VGG-16  & VGG-16 & VGG-16\\
			\hline
			Device &  XC7VX690T & Zynq ZCU102 & Zynq ZCU102 & XC7VX690T & Arria-10 GX1150  & Zynq ZCU102 & Zynq ZCU102 \\
			\hline
			Frequency(MHz) &  200 & 200 & 200 & 200 & 150  & 200 & 200\\
			\hline
			Precision &  16bit fixed & 16bit fixed & 16bit fixed & 8bit BFP & 16bit fixed  & 16bit fixed & 8bit int\\
			\hline
			Performance(GOP/s) &  270 & 476.7 & 223.4 & 281.5 & 232.2  & 495.4 & 990.8\\
			\hline
			DSP Efficiency(GOP/s/DSP) & 0.188 & 0.353 & 0.195 & 0.275 & 0.153  & 0.367 & 0.393\\
			\hline
			Logic cell Efficiency(GOP/s/K cells) & 0.577 & 1.222 & 0.405 & 1.213 & 1.442  & 1.270 & 2.446\\
			\hline
		\end{tabular}
	}
\end{table*}

\subsection{Resource utilization}

Table \ref{tab:breakdown} shows the resource utilization breakdown with the configuration ($N=48,M=28$). BRAMs are mainly used to construct the buffers and FIFOs (First In First Out). The parameter N determines the number of weight buffers in PUs and that of output FIFOs in DMA (Direct Memory Access). The product of parameters M and N determines the number of Psum buffers. Each DSP (Digital Signal Processing) can address a $16bit \times 16bit$ multiplcation or a $8bit \times 8bit$ multiplcation. The number of DSP can be calculated as $N \times M + 6$ when the width of operand is 16. The 6 DSPs are used to calculate the index address for WIB and the number of shift bit for VGM. Besides the buffers, the rest modules consume LUT and Flip-Flop (FF).

\begin{figure}
	\centering
	\includegraphics[scale=0.35]{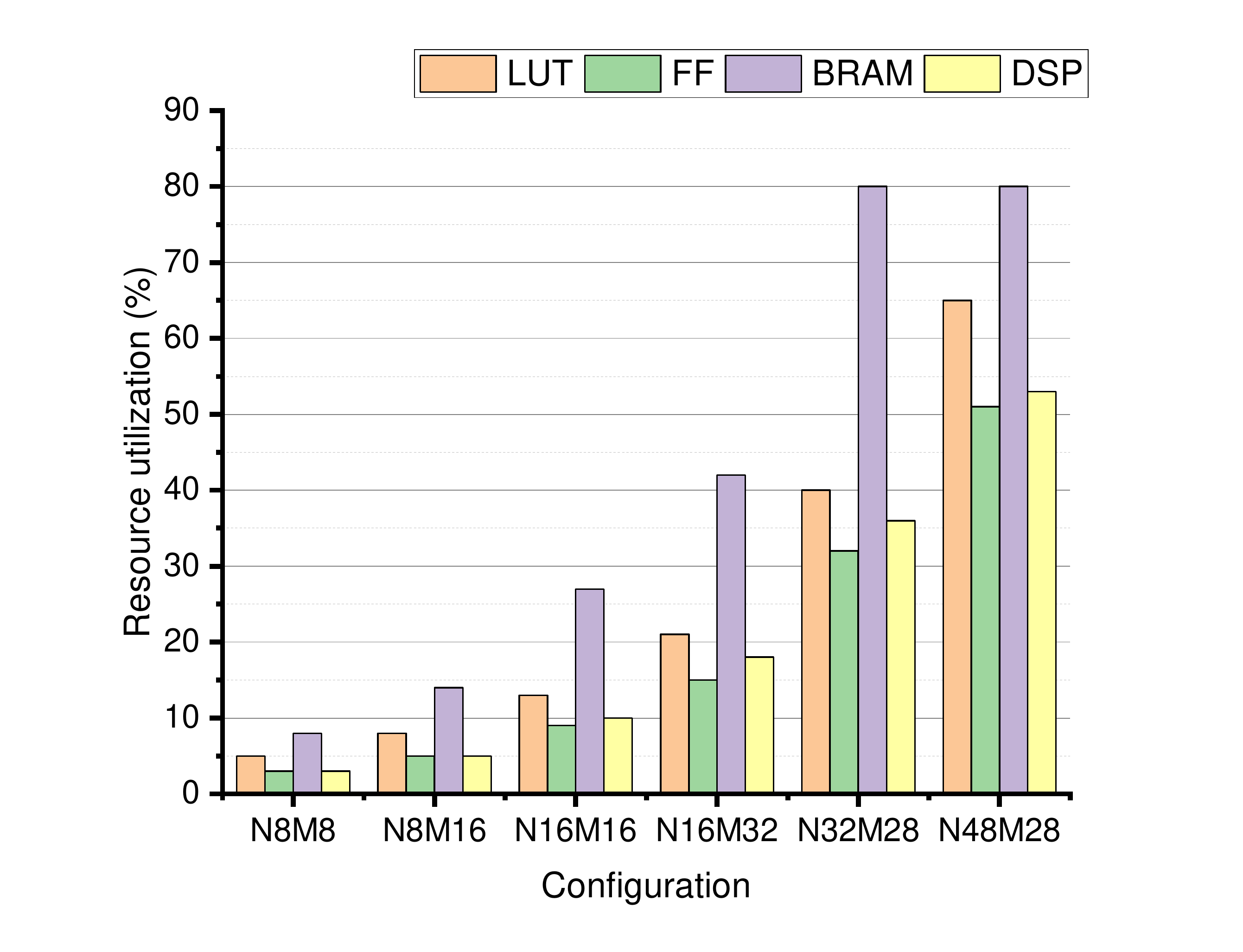}\\
	\caption{Resource utilization of the accelerator under different configuration.}\label{utilization}
\end{figure}

Fig. \ref{utilization} shows the resource utilization of different parallelism factors obtained from Xilinx Vivado tool (v.2017.2). The LUT utilization increases as the total number of PE $N\times M$ increases. The utilization of BRAM are mainly used to construct buffers and FIFOs.  When $(N,M)$ increases to a certain extent, some large FIFOs are implemented by LUTs and FFs rather than BRAM to meet the timing constraints.




\begin{figure}
	\centering
	\includegraphics[scale=0.30]{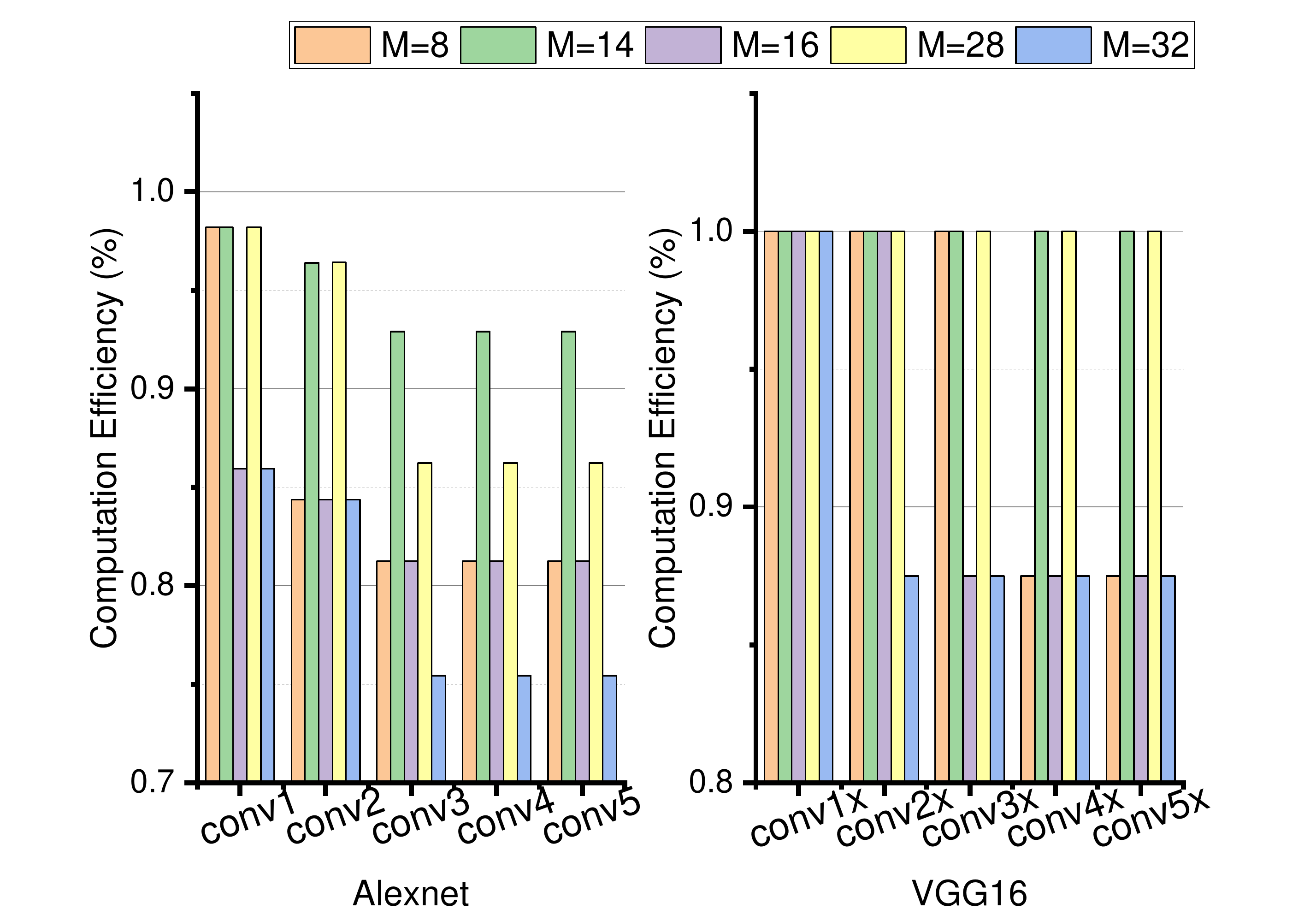}\\
	\caption{Computation efficiency of Alexnet and VGG-16 under different configuration.}\label{compute_efficiency}
\end{figure}

\begin{figure}
	\centering
	\includegraphics[scale=0.30]{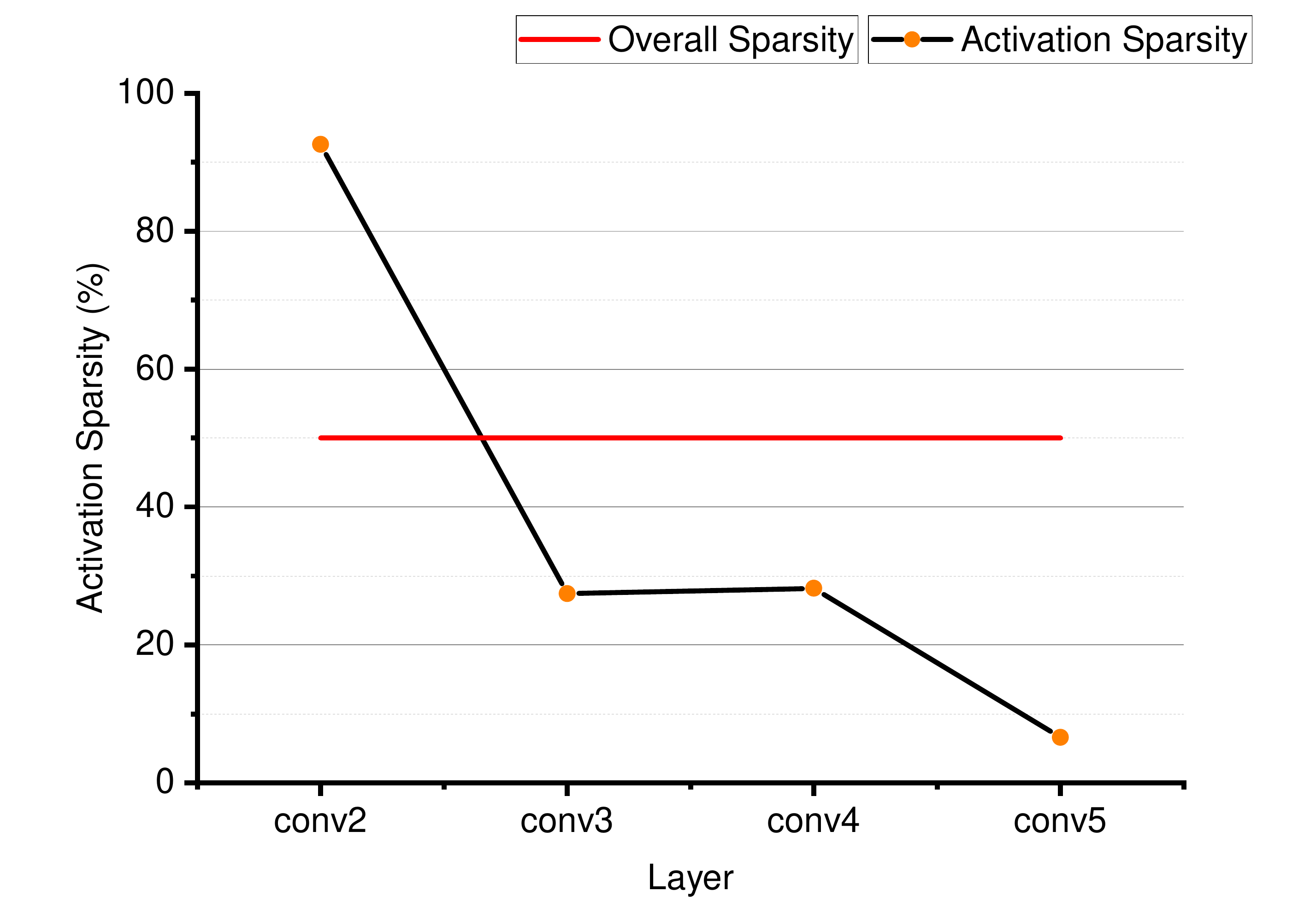}\\
	\caption{The activation sparsity of CONV layers in Alexnet. The overall sparsity is about 39.6\%.}\label{alexnet_sparsity}
\end{figure}

\begin{figure}
	\centering
	\includegraphics[scale=0.30]{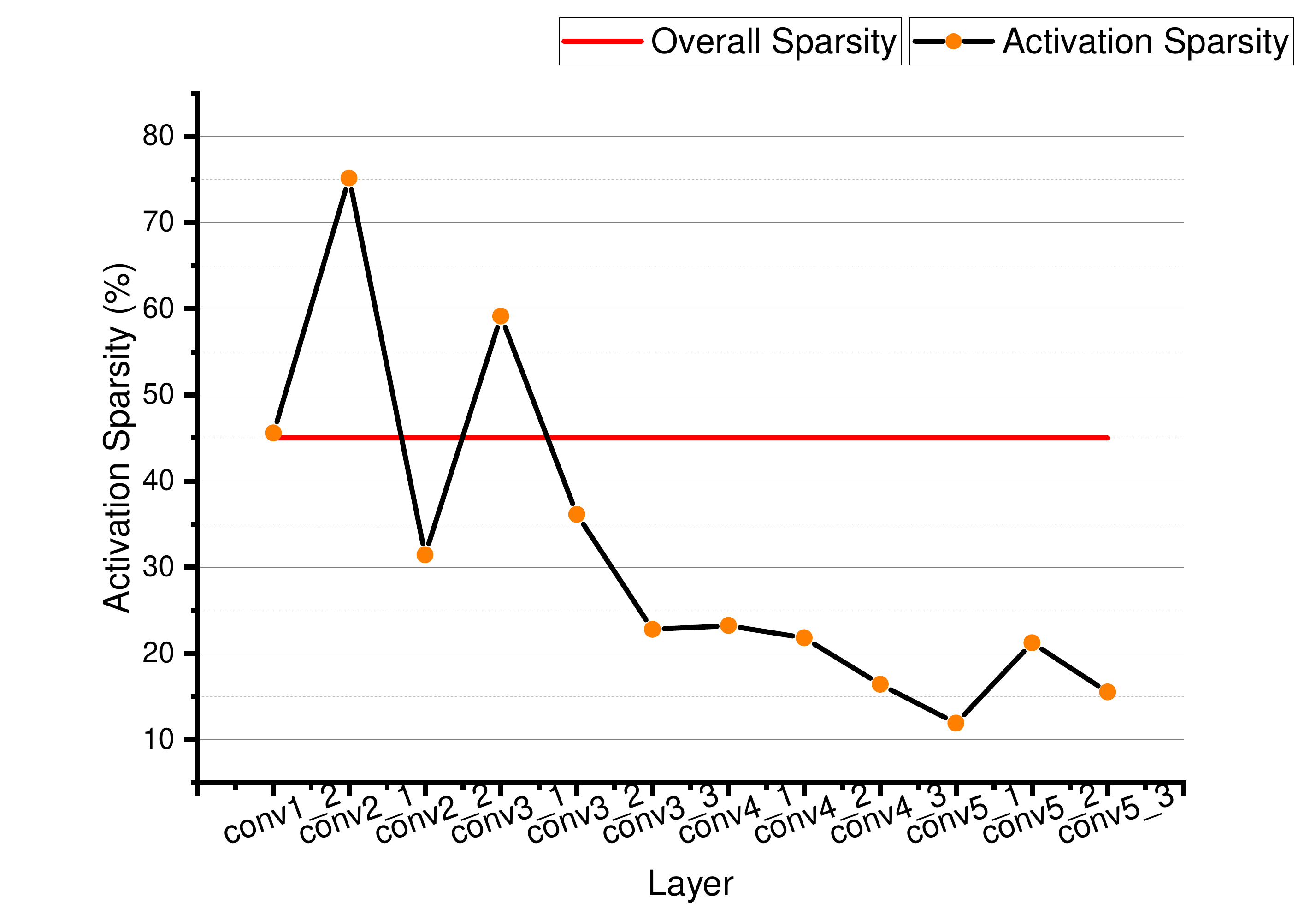}\\
	\caption{The activation sparsity of CONV layers in VGG16. The overall sparsity is about 39.5\%.}\label{vgg_sparsity}
\end{figure}

\begin{figure}
	\centering
	\includegraphics[scale=0.3]{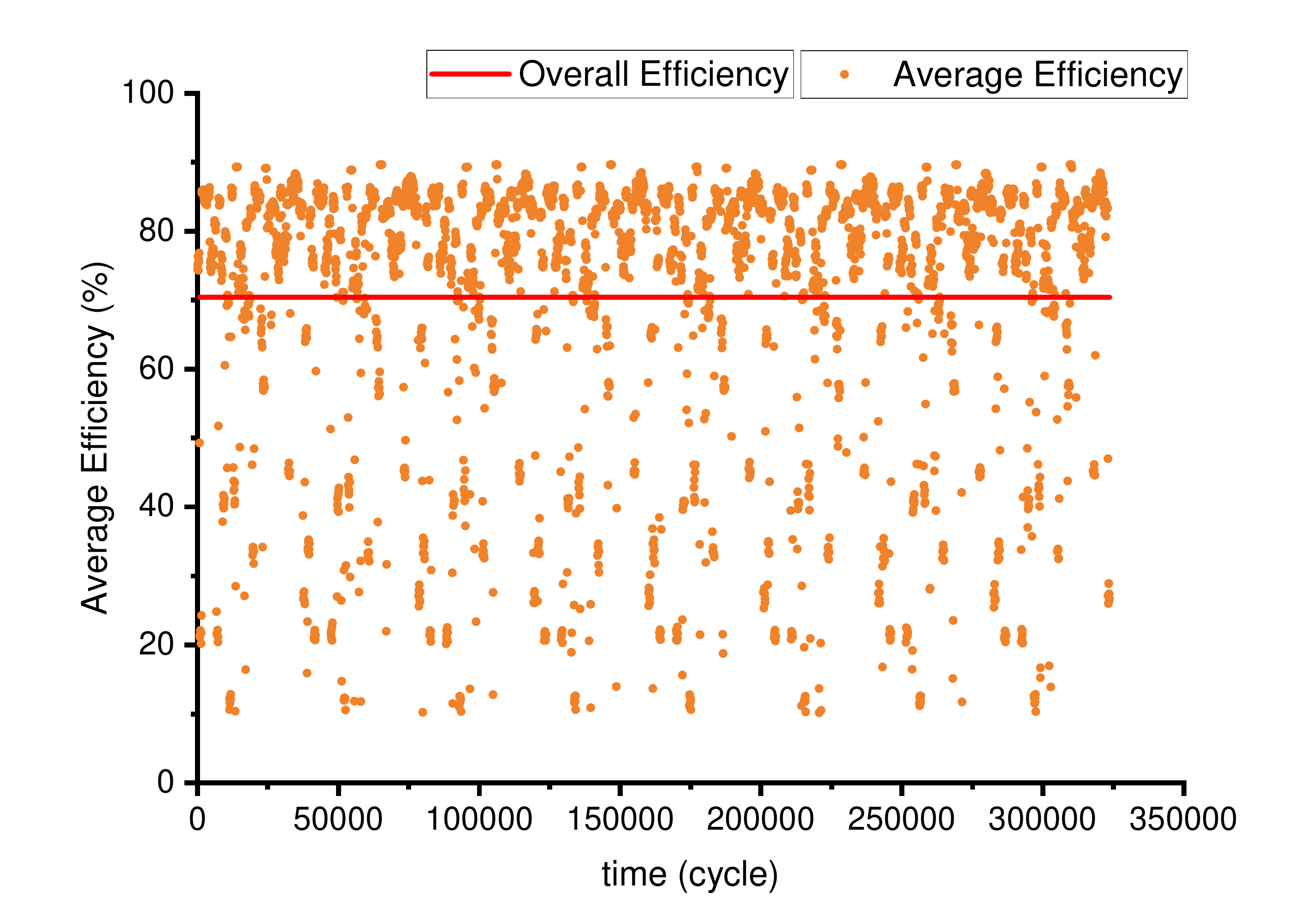}\\
	\caption{The proportion of active PEs on conv2\_1 of VGG-16. We sample the number of active PEs by simulation under the configuration of $<N,M>=<32,28>$. Each point represents the average efficiency of 100 samples.}\label{pe_statistic}
\end{figure}



\subsection{Computation Efficiency}
In this work, we do output channel parallelly across PUs. Benefiting from the shape-wise pruning, the load of each PU is balanced. When we map the network onto PEs, the inefficiency of our design mainly comes from two aspects: Dynamic Activation Inefficiency (DAI) and Dataflow Mapping Inefficiency (DMI). First, there are zero activations in input feature maps, which leads to some PEs gated. This pattern is designed to save energy deliberately. Second, the size of feature map cannot be divided by $M$ evenly, where $M$ is the number of PE in each PU. The DAI is dynamic and variable dependent on data sheet but irrespective of the proposed datafow whereas the DMI is dependent on the proposed dataflow and can be computed as the following equations.

We assume the size of a 3-D output feature map is $U \times V \times F$. According to our dataflow, when $V \textgreater M$, the average computation efficiency is shown as Eq(\ref{ce1}).
\begin{equation}\label{ce1}
Compute_{eff} = \frac{V}{\lceil V/M \rceil M}
\end{equation}
When $V \textless M$, the average compute efficiency is shown as Eq(\ref{ce2}).
\begin{equation}\label{ce2}
Compute_{eff} = \frac{U \times V}{\lceil \frac{U}{\lfloor \frac{M}{V} \rfloor} \rceil M}
\end{equation}

We measure the DMI on Alexnet and VGG-16. In Fig. \ref{compute_efficiency}, the computation efficiency which involves in DMI across different layers with different parallelism factors. Because both U and V in VGG-16 can be divided by 14, the computation efficiency keeps 100\% when M equals to 14 or 28. In conclusion, our sparse wise dataflow can maintain a high computation efficiency for different neuron networks. 

To analyse DAI, we firstly count sparse activations on convolutional layers of Alexnet and VGG-16 by using Pytorch vision. The dataset is ImageNet 2012. As shown in Fig. \ref{alexnet_sparsity}, layer conv5 shows the lowest activation sparsity below 10\%, and layer conv2 shows the highest activation sparsity over 90\%. The overall sparsity is about 39.6\%. As for VGG-16, Fig. \ref{vgg_sparsity} depicts that the last seven convolutional layers show a low sparsity below 30\% and layer conv2\_1 shows the highest sparsity about 75\%. In the mass, the overall activation sparsity is about 39.5\%. According to the activation sparsity, we can estimate the DAI.

Then we measure the DAI on conv2\_1 of VGG-16 by simulation. On each compute cycle, we sample the total number of active PEs and calculate the efficiency. As the number of total samples is too large, so we calculate an average efficiency every 100 samples. Fig. \ref{pe_statistic} shows the proportion of ungated PE. Indeed, part of PEs are gated to save energy. The proportion of active PE on conv2\_1 is positively related to the activation sparsity. The sample circuit is designed for simulation when the configuration parameter $<N,M>=<16,28>$, and does not be implemented on FPGA.

\subsection{Performance analysis}



\begin{figure}
	\centering
	\includegraphics[scale=0.3]{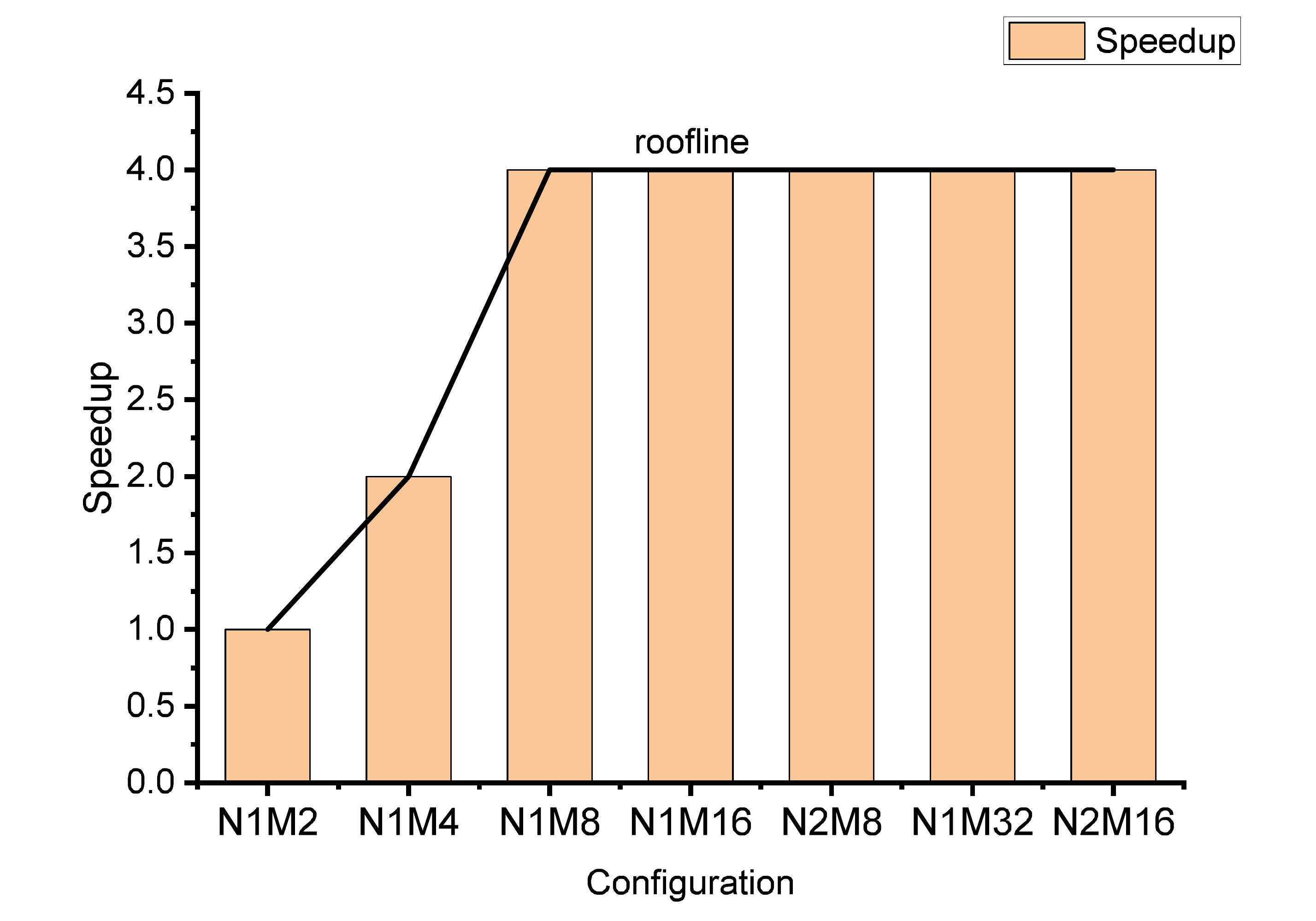}\\
	\caption{Roofline model for 16-bit fixed in FC layer under different configuration. When the number of PEs increases to 8, the performance touches the roof. Because the off-chip bandwidth is fully utilized.}\label{roofline_16bit}
\end{figure}

\begin{figure}
	\centering
	\includegraphics[scale=0.3]{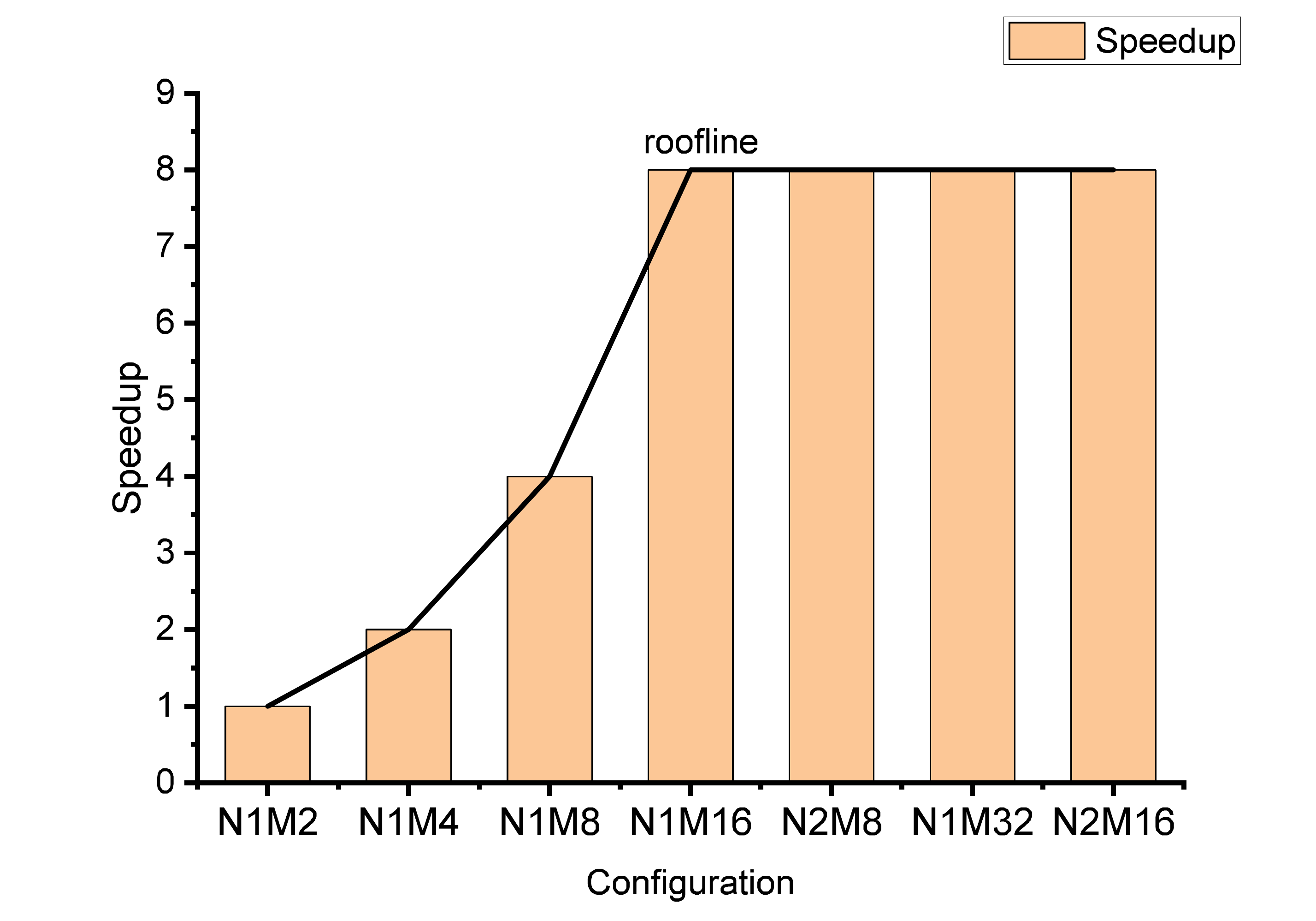}\\
	\caption{Roofline model for 8-bit int in FC layer under different configuration. The width of weight is 8-bit, so the performance touches the roof until the number of PEs increases to 16.}\label{roofline_8bit}
\end{figure}

In this section, we adopt the well-known roofline model\cite{zhang2015optimizing} for exploring the impact of insufficient off-chip bandwidth on performance. We set the bitwidth of AXI data bus which connects to DDR as 128-bit. First, we do not quantize the weight and change the configuration to find the optimal parameters of mapping the FC6 layer of VGG-16 onto our accelerator. We normalize all the performance number to that of "N1M2". As illustrated in Fig. \ref{roofline_16bit}, when the number of PE reaches to 8, the performance touches the roof. Second, we quantize the weights to 8-bit. We find that the performance touches the roof until the number of PE reaches to 16, and the roof becomes higher in Fig. \ref{roofline_8bit}. Because there is neither weight share nor weight reuse in FC layer. When the bitwidth of weight decreases by half, the number of transfered weights doubles in each DDR access. So the speedup also doubles. 


\begin{figure}
	\centering
	\includegraphics[scale=0.3]{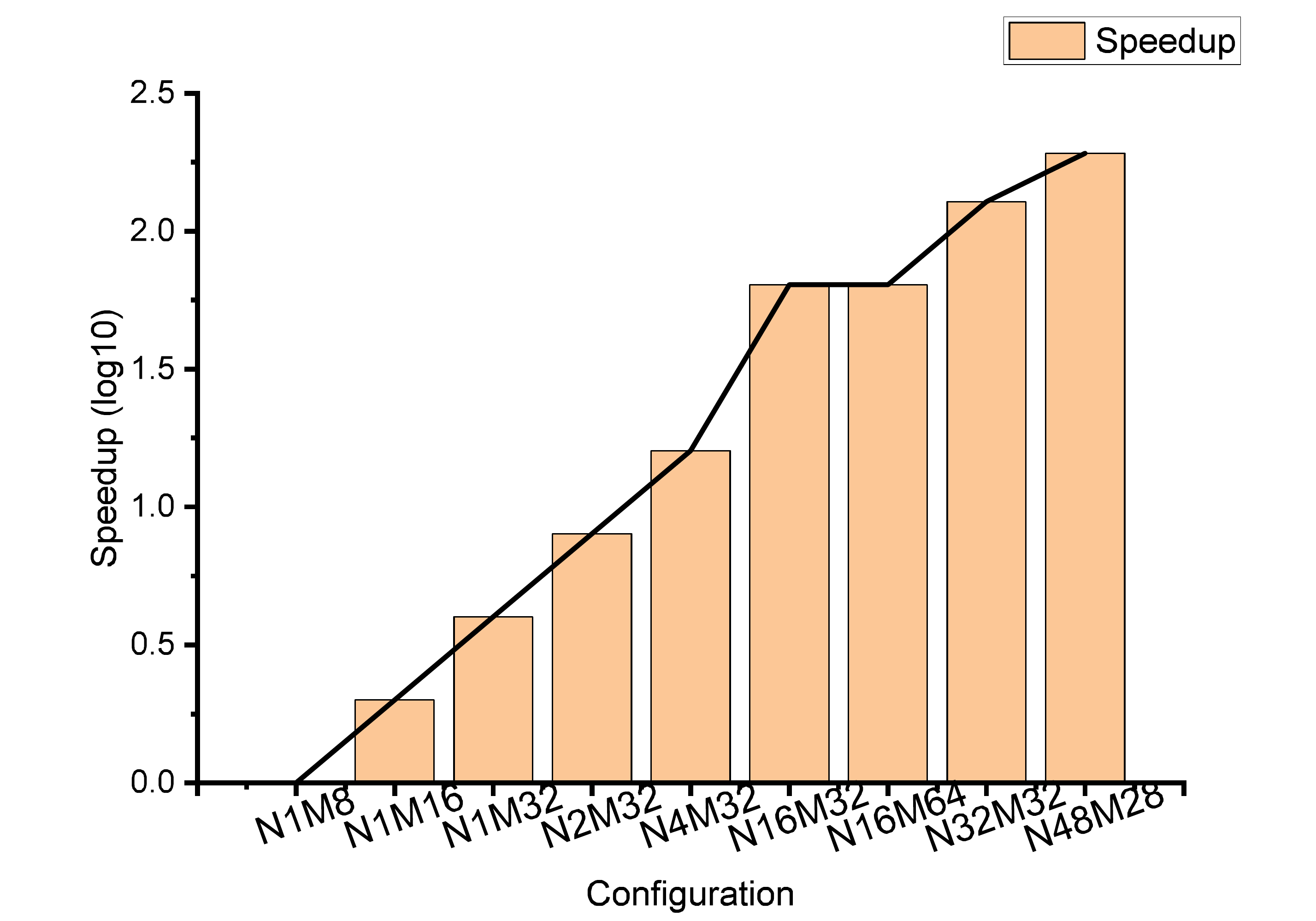}\\
	\caption{Roofline model for 16-bit fixed in CONV layer under different configuration. When the number of PEs in a PU increases from 32 to 64, the performance does not improve.}\label{roofline_conv}
\end{figure}

After that we map the conv2\_1 layer of VGG-16 onto our accelerator to explore the optimal parameters. To get a high compute efficiency, we set N as the divisor of the number of output channel F. We normalize all the performance number to that of "N1M8." In Fig.~\ref{roofline_conv}, configuration "N48M28" achieves the best peak performance.

Final, we analyze the performance of our implementation. We set the PE array size as $<N,M>=<48,28>$, which consists of 1344 PEs. In this configuration, the peak throughput can be calculated as $2 \times 0.2GHz \times 28 \times 48 = 537.6GOP/s$ when  the width of operand is 16. Specially, the proposed design supports to perform two $8bit \times 8bit$ multiplications in a PE with 2 DSPs, which leads to $1075.2\,GOP/s$ peak performance.

We compare our design with convolutional FPGA accelerators in Table \ref{tab:freq}. The performance in Table \ref{tab:freq} represents the effective throughput. \cite{ma2017optimizing, kala2019high} are dense CNN accelerators, and \cite{zeng2018framework, lu2019efficient} are sparse CNN accelerators. Both \cite{zeng2018framework} and \cite{lu2019efficient} only address the weight sparsity, but do not address the activation sparsity, so we compute our throughput with DMI which is definded in previous subsection. For the dense accelerator, the performance is computed by dividing the effective throughput with computation of dense network. According to Table \ref{tab:freq}, our accelerator achieves 987\,imag/s effective performance on structured sparse Alexnet which shows 1.5$\times$ to 6.7$\times$ speedup and 2.0$\times$ to 4.2$\times$ energy-efficiency compared with \cite{zeng2018framework,lu2019efficient,kala2019high}. As for VGG-16, our implementation achieves 48\,imag/s performance which is 2.2$\times$ to 2.3$\times$ speedup and 3.4$\times$ to 6.2$\times$ energy-efficiency compared with \cite{ma2017optimizing, lu2019efficient}. For the case of 8-bit int, we achieve 96\,imag/s performance which is 4.4$\times$ to 4.6$\times$ speedup and 6.1$\times$ to 11.2$\times$ energy-efficiency compared with \cite{ma2017optimizing, lu2019efficient}. The reason of the speedup is because our dataflow can effectively skip the sparse weight multiplications. In addition, this dataflow flexibly maps network onto PEs which leads to a high utilization of on-chip resources. Previous works cannot efficiently exploit the zeros or involve a low compute efficiency. Besides, we apply clock gate on unused PEs when the input activations equal to zero. So we get a higher energy-efficiency than previous works.

The proposed design achieves a higher performance of resource efficiency because we leverage the sparsity and achieve a high mapping efficiency, as tabulated in Table \ref{tab:per_eff}. The accelerator\cite{lu2019efficient} also leverages the sparsity but the DSP efficiency is encumbered by the low mapping efficiency due to the unbalanced load of PE. In addition, the logic cell efficiency is only 0.405\,GOP/s/K cells because the TLUT and CMUX\cite{lu2019efficient} consume a large number of logic resources. Reducing the bitwidth can help to improve the resource utilization efficiency. The design\cite{lian2019high} achieves a performance of 0.275\,GOP/s/DSP and 1.213\,GOP/s/K cells because it uses 8-bit block float point to represent activations and weights so that two multiplication operations can be carried out in a DSP slice. In the proposed design, if the width of data is 8-bit, the logic cell efficiency nearly improves 100\%.

\section{Conclusion}
In this work, we have proposed a sparse CNN FPGA accelerator with a sparse-wise dataflow to skip zero weights computations. Moreover we have exploited data statistics to minimize energy through zeros gating to avoid unnecessary computations. In addition, we have proposed a set of architecture optimization techniques for sparse CNNs. Experiments demonstrated that our implementation could achieve 987\,imag/s and 48\,imag/s performance for AlexNet and VGG-16 on Xilinx ZCU102, respectively, which provides 1.5$\times$ to 6.7$\times$ speedup and 2.0$\times$ to 6.2$\times$ energy-efficiency over previous CNN FPGA accelerators. Furthermore, the performance improvement and energy efficiency will be much larger if we achieve the sparsity of CNNs described in Cambricon-S\cite{zhou2018cambricon}.

\section{Acknowledgement}
This work is funded by the National Key R\&D Program of China (2018YFB0904902).

\ifCLASSOPTIONcaptionsoff
  \newpage
\fi



%
\bibliographystyle{IEEEtran}
\bibliography{literature}

%

\begin{IEEEbiography}[{\includegraphics[width=1in,height=1.25in,clip,keepaspectratio]{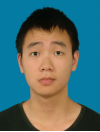}}]{Chaoyang Zhu} received the B.S. degree from College of Physics and Technology, Central China Normal University, China, in 2017. He is currently pursuing the master's degree in College of Information Science \& Electronic Engineering, Zhejiang University. His research interest includes hardware acceleration of neural network.
\end{IEEEbiography}

\begin{IEEEbiography}[{\includegraphics[width=1in,height=1.25in,clip,keepaspectratio]{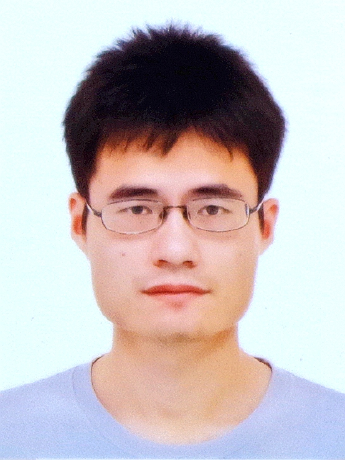}}]{Kejie~Huang}(M'13-SM'18) received his Ph.D degree from the Department of Electrical Engineering, National University of Singapore (NUS), Singapore, in 2014. He has been a principal investigator at College of Information Science Electronic Engineering, Zhejiang University (ZJU) since 2016. Prior to joining ZJU, he spent five years in the IC design industry including Samsung and Xilinx, two years in the Data Storage Institute, Agency for Science Technology and Research (A*STAR), and another three years in Singapore University of Technology and Design (SUTD), Singapore. He has authored or coauthored more than 30 scientific papers in international peer-reviewed journals and conference proceedings. He holds four granted international patents, and another eight pending ones.

His research interests include low power circuits and systems design using emerging non-volatile memories, architecture and circuit optimization for reconfigurable computing systems and neuromorphic systems, machine learning, and deep learning chip design. He is the Associate Editor of the IEEE TRANSACTIONS ON CIRCUITS AND SYSTEMS-PART \RNum{2}: EXPRESS BRIEFS.

\end{IEEEbiography}

\begin{IEEEbiography}[{\includegraphics[width=1in,height=1.25in,clip,keepaspectratio]{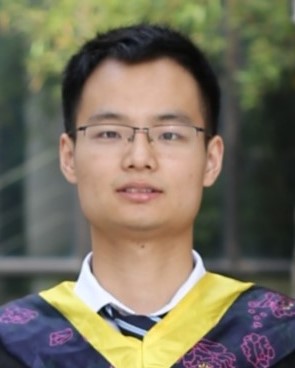}}]{Shuyuan~Yang} received the B.S. degree in electronic science and technology from Huazhong University of Science and Technology, Wuhan, China, in 2019. He is currently working toward the M.S. degree of electronic science and technology in Zhejiang University, Hangzhou, China. His current research interests include deep learning accelerator and network on chip.
\end{IEEEbiography}


\begin{IEEEbiography}[{\includegraphics[width=1in,height=1.25in,clip,keepaspectratio]{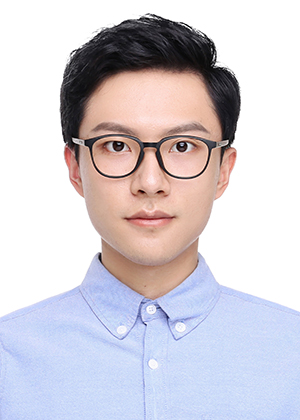}}]{Ziqi~Zhu} received the B.S. degree in Electronic and Information Engineering from the Zhejiang University, Hangzhou, China, in 2019. He is currently working towards the M.S. degree of Integrated Circuits. His current research interests include computer vision and 3D object detection.
\end{IEEEbiography}

\begin{IEEEbiography}[{\includegraphics[width=1in,height=1.25in,clip]{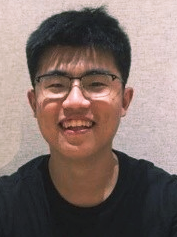}}]{Hejia~Zhang} received the B.E. degree in Bioengineering from Zhejiang University, Hangzhou, China, in 2017. He is currently working towards the Ph.D. degree in Computer Science at the University of Southern California, USA. His current research interests include robot learning from videos and submodular optimization for active learning.
\end{IEEEbiography}
\newpage

\begin{IEEEbiography}[{\includegraphics[width=1in,height=1.25in,clip,keepaspectratio]{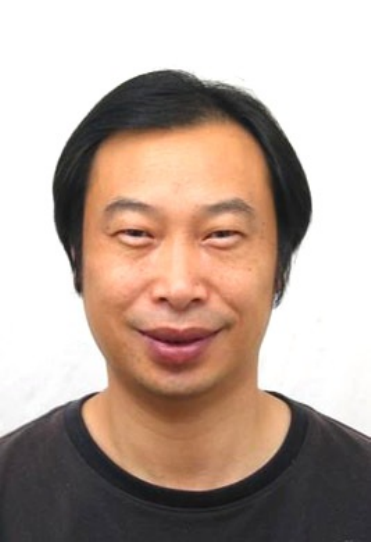}}]{Haibin~Shen}
	is currently a Professor with Zhejiang University, a member of the second level of 151 talents project of Zhejiang Province, and a member of the Key Team of Zhejiang Science and Technology Innovation. His research interests include learning algorithm, processor architecture, and modeling. His research achievement has been used by many major enterprises. He has published more than 100 papers on academic journals, and he has been granted more than 30 patents of invention. He was a recipient of the First Prize of Electronic Information Science and Technology Award from the Chinese Institute of Electronics, and has won a second prize at the provincial level.
\end{IEEEbiography}






\end{document}